\documentclass[prb,twocolumn,superscriptaddress,amsmath,amssymb,floatfix]{revtex4-2}
\usepackage{graphicx}
\usepackage{bm}
\usepackage[colorlinks=true,linkcolor=blue,citecolor=blue,urlcolor=blue]{hyperref}

\begin{document}

\title{Borrowed Planckian dissipation in magic-angle twisted graphene}

\author{Dino \v{S}krgi\'c}
\email{dinoskrg@buffalo.edu}
\affiliation{Department of Physics, University at Buffalo, State University of New York, Buffalo, New York 14260, USA}
\author{Dejan Stojkovic}
\affiliation{Department of Physics, University at Buffalo, State University of New York, Buffalo, New York 14260, USA}

\date{\today}

\begin{abstract}
In strange metals, the ratio of the coefficients producing signature linear-in-$T$ resistivity and the logarithmic specific heat  is free of the coupling between the electrons and whatever scatters them. In the materials studied so far that coupling is fixed by chemistry. Magic-angle twisted trilayer graphene provides a dial with a displacement field $D$ that breaks the mirror symmetry separating a light Dirac cone from the flat bands, and the Dirac carriers borrow the Planckian self-energy of the flat band with a weight proportional to $D^2+D_0^2$. Resistance curves taken at eight fields are consistent with the field acting on the scattering rate rather than on the number of carriers. The slope in temperature follows $a(1+D^2/D_0^2)$ without the quartic term that a depleting Dirac pocket adds, and the elastic part varies by less than 5\% between 0.20 and 0.30~V/nm while the inelastic part grows 2.5-fold up to 0.38~V/nm. The thermally averaged Kubo formula favors flat-band weight placed symmetrically about the Fermi level over a single level. Borrowed dissipation keeps the bath exponent and transport prefactor of the flat band's own electrons but carries exactly twice their entropy per unit of borrowed coupling whenever the bath has dynamics of its own, so the field-differential ratio is $2\pi^3/21\zeta(3)\approx2.46$ for a conformal flat band and, in a numerical search, at most 2.73 for monotonic bath spectra without a high-energy rate floor. The discriminating test is the entropy measured in steps of $D$.
\end{abstract}

\maketitle
\section{Introduction}
\begin{figure}[t]
\includegraphics[width=\columnwidth]{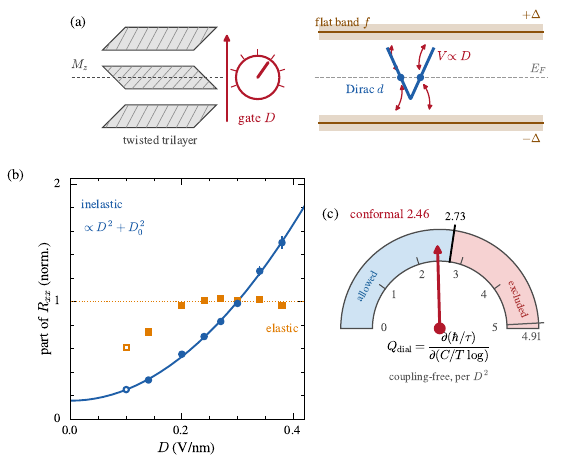}
\caption{Displacement field as a hybridization dial. (a) A displacement field $D$ breaking mirror symmetry $M_z$ of the alternating-angle twisted trilayer graphene, where the Dirac carriers at $E_F$ borrow flat-band character through $V\propto D$. (b) The two parts of the resistance of device D3 at $\nu=2.1$. Circles: the inelastic part, the 12--32~K slope ${\rm d}R_{xx}/{\rm d}T$, normalized. Squares: the elastic part $R_{\rm el}$ with the amplitude held to the $D^2$ law (free amplitudes: Fig.~\ref{fig:slopes}b), normalized to its mean for $D\geq0.20$~V/nm (dotted). Open symbols were not used in any fit. A field-dependent Dirac carrier number would bend the slope away from $D^2$. (c) The dial $Q_{\rm dial}$ of Eq.~(\ref{eq:Qdial}) is free of the coupling. A conformal flat band gives $2\pi^3/21\zeta(3)\approx2.46$ (needle), exactly half of the $4.91$ the same bath gives its own electrons.}
\label{fig:firstpage}
\end{figure}

A linear-in-$T$ resistivity at a rate close to the Planckian value, $\hbar/\tau\sim k_BT$, is shared by materials that otherwise have little in common~\cite{Bruin2013,Legros2019,HartnollMackenzie2022}. In hole-doped cuprates and heavy-fermion compounds it is paired with an electronic specific heat $C/T$ that rises logarithmically on cooling~\cite{Daou2009,Michon2019,Lohneysen2007,Custers2003}. These materials are chemically doped, so every added carrier brings a random potential with it. A prominent class of theories for these materials centers around this disorder. Specifically, it assumes there is a critical boson coupling to the electrons through spatially random vertices, giving a linear resistivity because the coupling is random~\cite{Esterlis2021,PatelGuoEsterlisSachdev2023,PatelLuntsSachdev2024,Li2024}. 
Magic-angle graphene, however, gets the same Planckian slope from the opposite end~\cite{Cao2020PRL,Jaoui2022,ZhangLi2026}. Its carriers are added by a gate, there are no dopant atoms, and the main inhomogeneity is a slow variation of the twist angle~\cite{Uri2020,Wilson2020}, intuitively a `clean' version of the disordered strange metal. Whether its linear resistivity is electronic or comes from phonons is still debated~\cite{Polshyn2019,WuHwangDasSarma2019,Shilov2026}, and no logarithm in $C/T$ has been isolated yet.

The clean and the disordered strange metals have therefore been compared only through the Planckian coefficient, and every theory of it carries a coupling that is not known independently~\cite{Varma1989,SachdevYe1993,Kitaev2015,SongJianBalents2017,Chowdhury2018,ChowdhuryRMP2022,FaulknerLiuMcGreevyVegh2011,SachdevReview2025}. Even within the cuprates, transport and thermodynamics have been compared at a common, empirically fitted coupling~\cite{Varma2020,Michon2023,Li2024}. This coupling can drop for a specific combination of the coefficients. Specifically, for electrons scattered by a local critical bath, the slope $\alpha$ of the transport rate, $\hbar/\tau_{\rm tr}=\alpha k_BT$, and the coefficient $B$ of the logarithm in $C/T$ form (with the band Sommerfeld coefficient $\gamma_b$) a ratio $Q=\alpha\gamma_b/B$ that depends only on the bath~\cite{SkrgicStojkovic2026}. It is fixed by how slowly the bath relaxes, an exponent $\beta$ seen in photoemission and optics, and by the share $\eta$ of the critical entropy that relaxes current.

A metal that is chemically doped is therefore classified at a single point, because its coupling, band factor and the split of its entropy between bath and electrons all come with the compound, and doping or pressure changes the bath along with the coupling. Magic-angle graphene has not entered it at all, since its logarithm is hidden. What is missing on both sides is a system in which the coupling between itinerant carriers and a fixed Planckian sector can be turned by hand. That raises two questions: whether such a dial exists, and what a coupling-free ratio measures when the carriers borrow their dissipation from a neighbor instead of creating it themselves.

We answer the first question with alternating-twist trilayer graphene near the magic angle (MATTG). At zero displacement field it splits into a mirror-even sector that copies magic-angle bilayer graphene and a mirror-odd sector that is a graphene-like Dirac cone~\cite{Khalaf2019,Carr2020}, and while the top and bottom layers are equivalent the two do not talk to each other. They become linearly coupled through the displacement field, $D$~\cite{Khalaf2019,Lei2021,Christos2022,Phong2021}. In the heavy-fermion description of moir\'e graphene the flat bands supply localized $f$ electrons and the Dirac cone supplies itinerant carriers~\cite{SongBernevig2022,HuBernevigTsvelik2023,ChouDasSarma2023,LauColeman2025}. This makes $D$ a gate-tunable $f$--$c$ hybridization~\cite{RamiresLado2021,Yu2023THF} (Fig.~\ref{fig:firstpage}a). The same gate tunes the superconductivity of MATTG~\cite{Park2021,Hao2021,Kim2022,Zhou2025}, and transport, compressibility and Dirac spectroscopy trace a $D$-dependent phase diagram in heavy-fermion terms~\cite{Zhang2026PRL,Pierce2025,Shen2023}. However, there is a competing framework. In Ref.~\cite{LiangZhaoZhang2026} $D$ mainly shifts the Dirac cone, by an amount of order $\nu(M_1D)^2/U$, and moves carriers into the bilayer sector, while a momentum-selective Mott gap suppresses the effective hybridization.

A measurement that can decide between the readings is the normal-state resistance, already measured on a single device~\cite{Zhang2026PRL}. There, eight $R_{xx}(T)$ curves of one device at filling $\nu=2.1$ for $D=0.10$--$0.38$~V/nm have been reported. Each rises from 12~K, bends over and peaks broadly near 57~K, which the authors attribute to Kondo coherence. The rise grows about fourfold over the set while the peak stays between 58 and 61~K from 0.20~V/nm up. A change in the number of Dirac carriers rescales the Matthiessen sum by a factor that depends on $D$, which bends the temperature-linear slope away from $D^2$ and, for charged scatterers, moves the elastic part as well, whereas a field that acts on a scattering rate adds $D^2$ to the inelastic part alone. Read this way, the curves are consistent with the second behavior (Fig.~\ref{fig:firstpage}b). The simplest theory that produces it also fixes the temperature profile and favors flat-band weight placed symmetrically about the Fermi level.

We answer the second question too. The hybridization that lets the Dirac carriers scatter off the flat band also carries flat-band entropy into them, and differences between fields remove the coupling together with every field-independent background. At second order in the hybridization, the field-differential ratio is set solely by the flat band's bath. Borrowed dissipation inherits the bath exponent and the transport prefactor of the flat band's own electrons unchanged, and for any bath with dynamics of its own it carries exactly twice their entropy per unit of coupling. A field damped by the electrons it scatters breaks this rule, and the field-induced specific heat then stops being logarithmic. The trilayer graphene thus introduces a knob into the strange-metal classification of Ref.~\cite{SkrgicStojkovic2026}, which cuprates and heavy fermions lack, and the shape of the entropy in steps of $D$ identifies the bath class (Fig.~\ref{fig:firstpage}c).

\section{The Model}\label{sec:model}

Let $d_{\mathbf k}$ annihilate a Dirac electron of the mirror-odd sector with dispersion $\xi_d(\mathbf k)$, and let $f$ denote the electrons of the mirror-even flat bands. Their Green's function at the Dirac momentum, $G_f(\omega,T)$, already contains their own interactions. The hybridization is
\begin{equation}
\begin{split}
H_{\rm hyb}&=\int d^2r\,\big[V(\mathbf r)\,d^\dagger(\mathbf r)f(\mathbf r)+{\rm h.c.}\big],\\
V(\mathbf r)&=v_DD+h(\mathbf r).
\end{split}
\label{eq:V}
\end{equation}
The mirror $M_z$ sends $d\to-d$, $f\to f$ and $D\to-D$, so the coefficient of $d^\dagger f$ must be mirror-odd. The displacement field supplies a uniform one, $v_DD$~\cite{Khalaf2019,Lei2021,Christos2022}. Any other mirror breaking, from unequal twist angles~\cite{Turkel2022} or an asymmetric dielectric environment, supplies a quenched piece $h(\mathbf r)$ with $\langle h\rangle=0$ and $\langle|h|^2\rangle=s^2$. A built-in uniform mirror breaking would instead add a constant to $v_DD$ and shift the minimum of every observable away from $D=0$. In the continuum model $V$ is a matrix in sublattice and valley space. Here, we use the scalar which is its projection onto the Dirac states at the Fermi level. In the heavy-fermion description the field also couples the Dirac cone to the dispersive $c$ electrons~\cite{Yu2023THF}; that coupling reshapes the Dirac dispersion and so enters through the carrier factor $\mathcal K$ of Sec.~\ref{sec:transport}, the channel tested in Sec.~\ref{sec:data}.

We assume local flat band self-energy, as done in dynamical mean-field and large-$N$ treatments~\cite{Georges1996,PJF1995}. The Dirac carriers thus dominate the conductance, since their velocity exceeds that of the flat bands by two orders of magnitude, and the experiment confirms it: at $\nu=6$, where the flat bands are full, $R_{xx}$ is flat in temperature at every $D$~\cite{Zhang2026PRL}. A parallel flat-band channel carries a fraction $\sim(N_f/N_d)(v_f/v_d)^2(\tau_f/\tau_d)\sim10^{-3}$--$10^{-2}$ of the current, so its own temperature dependence enters below a percent of $R$. Coupling a critical fermion system to a reservoir modifies the dynamics, such as Yukawa--SYK metal relaxing into a cold bath~\cite{Hosseinabadi2023}. Here the Dirac carriers are the reservoir, and their effect on the flat band is cut off by their small density of states. The feedback parameter is $\langle|V|^2\rangle N_d/\Gamma_f$, where $N_d$ is the Dirac density of states per moir\'e cell and $\Gamma_f$ is the flat-band width.

Integrating out $f$ gives the Dirac self-energy
\begin{equation}
\Sigma_d(\mathbf r,\omega)=|V(\mathbf r)|^2\,G_f(\omega,T)\big[1+O(\langle|V|^2\rangle N_d/\Gamma_f)\big].
\label{eq:sigma}
\end{equation}
This is the structure of the periodic Anderson model, whose conduction-electron Green's function is exactly $[\omega-\xi-V^2/(\omega-\varepsilon_f-\Sigma_f)]^{-1}$ with $\Sigma_f$ the full $f$ self-energy~\cite{Hewson1993,Georges1996}. The only approximation is that $\Sigma_f$ is evaluated at $V=0$, the self-energy of just the flat band. The corrections, from the change of $G_f$ by the Dirac electrons and from the connected four-point function of the interacting $f$ electrons, each carry an internal Dirac line and are suppressed by the feedback parameter. Averaging over $h$ replaces $|V|^2$ by
\begin{equation}
\langle|V|^2\rangle=v_D^2\big(D^2+D_0^2\big),\qquad D_0=s/v_D,
\label{eq:V2}
\end{equation}
since the cross term vanishes for $\langle h\rangle=0$. Replacing the local coupling by its average in the self-energy requires $h$ to vary on scales short compared with the Dirac mean free path, $\ell\approx\hbar v/\Gamma\sim100$~nm, with $\Gamma$ the total Dirac rate. We consider the smooth limit in Sec.~\ref{sec:background}. The field-differential quantities below depend only on the uniform part $v_DD$, since the cross term vanishes, so $h$ enters only the zero-field background. The rate borrowed in this way relaxes current because the flat band's own scattering does, through moir\'e umklapp and disorder, which is what a local $\Sigma_f$ expresses.

Near $\nu=2$ the flat-band weight at the Dirac momentum is split into lower and upper Hubbard bands. We place it at $\pm\Delta$ with equal weight, $G_f=\frac12\sum_{s=\pm}[\omega-s\Delta-\Sigma_f(\omega)]^{-1}$, and compare with a single level at $+\Delta$ in Sec.~\ref{sec:background}. For $|\omega|,|\Sigma_f|\ll\Delta$ the symmetric case gives (Appendix~\ref{app:expand})
\begin{equation}
\Sigma_d(\omega)=-Y\big[\omega-\Sigma_f(\omega)\big]+O(\Delta^{-4}),\qquad Y=\frac{\langle|V|^2\rangle}{\Delta^2}.
\label{eq:lowE}
\end{equation}
The Dirac carriers inherit the flat band's self-energy with weight $Y$, together with a static mass renormalization. The two Hubbard bands push the Dirac level in opposite directions, so there is no level shift at this order.

The flat band itself is taken to be unchanged by the field, although the superconducting and correlated states of MATTG depend strongly on $D$~\cite{Park2021,Hao2021,Kim2022,Zhou2025}. By mirror symmetry the field reaches the mirror-even flat band only at order $D^2$, through the hybridization (with a small feedback), or through remote bands. The ordered states are set by scales of order 0.1~meV and respond to percent-level changes of the flat band, as well as to the Dirac admixture itself, which opens a pairing and pair-breaking channel, whereas the normal-state profile probes the Hubbard splitting of about 20~meV found in Sec.~\ref{sec:background}. The resistance maximum provides the check. At $\nu=2.1$ it stays within 3~K from 0.20 to 0.38~V/nm while the amplitude grows 2.5-fold, and once the background of Sec.~\ref{sec:background} is included a single $T^*$ fits all eight fields, so $\Delta/\kappa$ changes by no more than about 5\%.

\section{Transport through the dial}\label{sec:transport}

For a momentum-independent self-energy the current vertex is not renormalized, and the Kubo bubble gives (Appendix~\ref{app:kubo})
\begin{equation}
\rho(T,D)=\mathcal K\left[\int d\omega\left(-\frac{\partial n_F}{\partial\omega}\right)\frac{1}{\Gamma(\omega,T)}\right]^{-1},
\label{eq:rho}
\end{equation}
with $\Gamma=\Gamma_c+\langle|V|^2\rangle\,g(\omega,T)$, $g=-{\rm Im}\,G_f$, $\Gamma_c$ the elastic rate of the Dirac carriers, and $\mathcal K$ a constant set by the Dirac density and velocity. The real part of $\Sigma_d$ drops out when the Dirac density of states varies little over $\Gamma$, which is why local self-energies give resistivities set by ${\rm Im}\,\Sigma$~\cite{PJF1995,Georges1996}.

If $g$ varies little over $|\omega|\lesssim k_BT$ the thermal average is trivial and
\begin{equation}
R(T,D)=R_{\rm el}+W(D)\,F(T),\qquad W\propto D^2+D_0^2,
\label{eq:factor}
\end{equation}
with one profile $F\propto g(0,T)$ at all fields and an elastic part $R_{\rm el}\propto\Gamma_c$. For a Planckian flat-band width $\Gamma_f=\kappa k_BT$ at $\omega=0$,
\begin{equation}
F(T)=\frac{x}{1+x^2},\qquad x=\frac{T}{T^*},\qquad k_BT^*=\frac{\Delta}{\kappa},
\label{eq:F}
\end{equation}
for either level structure. The profile rises linearly for $T\ll T^*$, peaks at $W/2$ at a temperature that does not depend on $D$, and falls as $T^*/T$ above it. Its linear rise is the Lorentzian tail of flat-band weight of Planckian width, seen through a window whose size the field sets.

In the data window, $T^*/5\lesssim T\lesssim T^*$, the relevant frequencies are $\omega/\Delta\sim x/\kappa$, so the frequency dependence of $g$ matters unless $\kappa$ is large. Equation~(\ref{eq:rho}) is a harmonic mean, which is not linear in $\langle|V|^2\rangle$ when $g$ depends on $\omega$. It generates a $D^4$ term and moves the maximum away from $\Delta/\kappa$. We evaluated Eq.~(\ref{eq:rho}) exactly with $-{\rm Im}\,\Sigma_f=\kappa[\omega^2+(\pi k_BT)^2]^{1/2}/\pi$, with and without its marginal real part, on synthetic curves built to resemble the data (Table~\ref{tab:synthetic}). The quadratic law holds to better than 1\% for symmetric weight with $\kappa\gtrsim1$ and for a single level with $\kappa\gtrsim2$. A single level with $\kappa\le1$ gives a quartic share of $-6$ to $-34\%$ and moves the maximum to 24--40~K.

\begin{table}[t]
\caption{Exact thermal average of Eq.~(\ref{eq:rho}) on synthetic curves with $W=17.4(D^2+0.16^2)$~k$\Omega$, $R_{\rm el}=0.6$~k$\Omega$ and $\Delta=\kappa k_B\times57.8$~K, analysed like the data. Quartic share: the $D^4$ term of a fit $a+cD^2+eD^4$ to the 12--32~K slopes, at $D=0.38$~V/nm. $T_{\max}$: position of the resistance maximum.}
\label{tab:synthetic}
\begin{ruledtabular}
\begin{tabular}{lccc}
flat-band level(s) & $\kappa$ & quartic & $T_{\max}$ (K)\\
\hline
single, marginal ${\rm Re}\,\Sigma_f$ & 0.5 & $-34\%$ & 24\\
single, no ${\rm Re}\,\Sigma_f$ & 1 & $-14\%$ & 31\\
single, marginal ${\rm Re}\,\Sigma_f$ & 1 & $-6.5\%$ & 40\\
symmetric $\pm\Delta$ & 1 & $-0.6\%$ & 29\\
single, marginal ${\rm Re}\,\Sigma_f$ & 2 & $+0.5\%$ & 50\\
symmetric $\pm\Delta$ & 2 & $+0.1\%$ & 40\\
\end{tabular}
\end{ruledtabular}
\end{table}

The real part of Eq.~(\ref{eq:lowE}) makes the Dirac carriers heavier. A Planckian flat-band width comes with a marginal real part, $\partial_\omega{\rm Re}\,\Sigma_f=-\lambda_f\ln(\omega_c/\max(|\omega|,k_BT))$, by the Kramers--Kronig relations, so that for $T\ll T^*$
\begin{equation}
\frac{m^*}{m}=1+Y\big[1+\lambda_f\ln(\omega_c/k_BT)\big],\qquad Y\propto D^2+D_0^2 .
\label{eq:mass}
\end{equation}
The mass acquires a static part and a logarithm inherited from the flat band, both scaled by the same $D^2+D_0^2$ as the resistance. A flat band whose width is linear in $T$ but has no marginal real part, a quasi-elastic bath, would give only the static part. The logarithm in the mass has a counterpart in the entropy, which Sec.~\ref{sec:entropy} computes.

\section{Results}\label{sec:data}

We digitized the $R_{xx}(T)$ curves of device D3 ($\theta=1.27^\circ$, $\nu=2.1$) from Supplemental Fig.~S15b of Ref.~\cite{Zhang2026PRL} at $D=0.10$, 0.14, 0.20, 0.24, 0.27, 0.30, 0.34 and 0.38~V/nm, colour-tracking the native raster at 600~dpi, removing outliers by iterative robust smoothing and checking the axis calibration against the tick marks. Two independent digitizations agree to 0.2--0.6~$\Omega$/K in slope. Below 12~K the traces at 0.10, 0.14 and 0.38~V/nm overlap their neighbours, so all fits use 12--78~K, and the $D=0.10$ curve was kept out of every fit until the tests below. Only ratios and differences of resistances enter, so the unknown geometric factor drops out. Device D3 lies below the magic angle of the trilayer, so its flat bands are wider than near $1.5^\circ$; the analysis needs only that they carry a Planckian width and a Hubbard splitting at $\nu=2.1$, which the fitted $\kappa$ and $\Delta$ of Sec.~\ref{sec:background} test. Scripts and extracted points are available at~\cite{SkrgicMATTGcode}.

\begin{figure*}[t]
\includegraphics[width=0.95\textwidth]{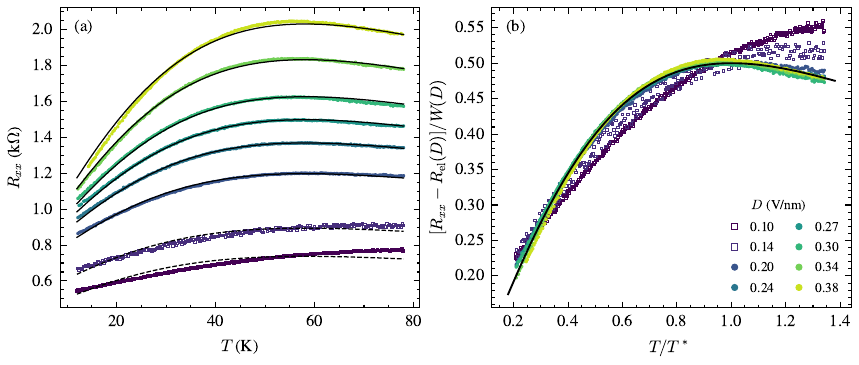}
\caption{(a) $R_{xx}(T)$ of MATTG device D3 at $\nu=2.1$, digitized from Supplemental Fig.~S15b of Ref.~\cite{Zhang2026PRL}. Solid lines: Eq.~(\ref{eq:factor}) with the profile (\ref{eq:F}) fitted to the five curves at $D\geq0.20$~V/nm with four shared parameters ($R_{\rm el}$, $W_2$, $D_0$, $T^*$), $W=W_2(D^2+D_0^2)$. Dashed: the same law extrapolated to $D=0.14$ and 0.10~V/nm (open symbols), with only the elastic offset adjusted. (b) $[R_{xx}-R_{\rm el}(D)]/W(D)$ against $T/T^*$, with $W(D)$ from the law, not fitted per curve; black, $x/(1+x^2)$. The two lowest-field curves follow the profile up to $T^*$ and rise above it, as a small field-independent background would make them (Sec.~\ref{sec:background}).}
\label{fig:collapse}
\end{figure*}

The matrix of temperature-dependent parts, $R(T_i,D_j)-R(12\,{\rm K},D_j)$, is rank one to 99.9\% over 14--78~K, through the maximum, with a second singular value 2--3\% of the first depending on smoothing, comparable to the digitization noise. A local quadratic places the maximum at 58--61~K for every field from 0.20 to 0.38~V/nm while the amplitude grows 2.5-fold; at 0.14~V/nm it lies at 71~K, which the field-independent background of Sec.~\ref{sec:background} accounts for. With a free offset and amplitude per curve and one shared scale, the profile of Eq.~(\ref{eq:F}) fits the seven curves at $D\geq0.14$~V/nm with an rms residual of 8.0~$\Omega$ and $T^*=57.8$~K, close to the 7.5~$\Omega$ of a shared sixth-order polynomial (Fig.~\ref{fig:collapse}). This establishes the factorization, though not the Lorentzian form in particular. Other one-parameter peaked profiles do about as well (Table~\ref{tab:shapes}); the data ask for a peak near 57~K with the curvature of $x/(1+x^2)$ or $xe^{-x}$, and they penalize broader profiles such as $x/(1+x)^2$ or $\ln(1+x)/(1+x)$ and reject a monotonic saturating one. Letting the width grow as $T^n$ gives $n=1.06$--$1.19$ across fit windows and with or without that background, $n=1.1\pm0.1$.

\begin{table}[t]
\caption{Shared one-parameter profiles $F(T/T_s)$ fitted to the seven curves at $D\geq0.14$~V/nm, 12--78~K, with a free offset and amplitude per curve (15 parameters). A shared sixth-order polynomial profile gives 7.5~$\Omega$.}
\label{tab:shapes}
\begin{ruledtabular}
\begin{tabular}{lcc}
profile & $T_s$ (K) & rms ($\Omega$)\\
\hline
$\tanh(x)/(1+x^2)^{1/2}$ & 54.9 & 7.7\\
$x/(1+x^2)$, Eq.~(\ref{eq:F}) & 57.8 & 8.0\\
$x\,e^{-x}$ & 58.3 & 8.5\\
$x/(1+x^2)^{3/4}$ & 42.4 & 11.0\\
$x/(1+x)^2$ & 61.6 & 14.0\\
$\ln(1+x)/(1+x)$ & 36.8 & 15.9\\
$x/(1+x)$ (no maximum) & 1 & 31.4\\
\end{tabular}
\end{ruledtabular}
\end{table}

\begin{figure*}[t]
\includegraphics[width=0.95\textwidth]{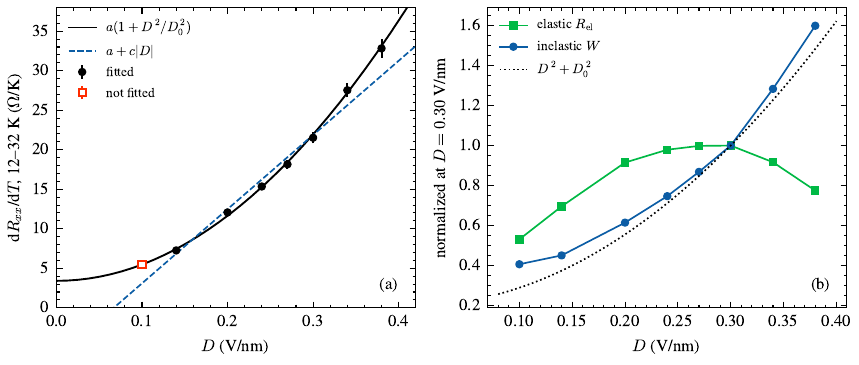}
\caption{(a) Slope ${\rm d}R_{xx}/{\rm d}T$ over 12--32~K versus $D$. Filled circles were fitted; the open square at $D=0.10$~V/nm was not. Error bars combine the fit scatter and the spread over fit windows. Solid: $a(1+D^2/D_0^2)$. Dashed: the best linear law. (b) Elastic offset $R_{\rm el}$ (squares) and inelastic amplitude $W$ (circles), both left free for each curve with the profile of Eq.~(\ref{eq:F}) at a shared $T^*=57.8$~K, normalized at $D=0.30$~V/nm. Dotted: $D^2+D_0^2$ with $D_0=0.15$~V/nm, not fitted.}
\label{fig:slopes}
\end{figure*}

The slope ${\rm d}R/{\rm d}T$ over 12--32~K measures the amplitude without reference to any profile. With errors that combine the fit scatter and the spread over fit windows between 12--30 and 14--34~K, it follows $a(1+D^2/D_0^2)$ with $a=3.4\pm0.3$~$\Omega$/K, $D_0=0.129\pm0.007$~V/nm and $\chi^2=2.9$ for 5 degrees of freedom (Fig.~\ref{fig:slopes}a); the window spread is correlated between fields, so this is not a strict goodness of fit. A free exponent in $a+cD^p$ gives $p=1.85\pm0.18$, and a quartic term contributes $-6\pm10\%$ at 0.38~V/nm. A linear law $a+c|D|$ gives $\chi^2=23$ and needs a negative zero-field slope, $a=-6.3$~$\Omega$/K. With its parameters fixed by the seven curves, the quadratic law predicts 5.47~$\Omega$/K at $D=0.10$~V/nm, where the curve gives $5.48\pm0.6$~$\Omega$/K and the linear law 3.1~$\Omega$/K. Mirror symmetry makes every smooth observable start as $a+cD^2$, so the leading power is expected. What the data add is that the quadratic law holds over a 4.5-fold change with no measurable quartic term.

For $D\geq0.20$~V/nm the elastic part depends little on the field, as Eq.~(\ref{eq:factor}) requires. With the four-parameter law $R=R_{\rm el}+W_2(D^2+D_0^2)F(T/T^*)$ fitted to those five curves, a single shared $R_{\rm el}=0.633$~k$\Omega$ gives 17~$\Omega$ rms and separate offsets give 6.6~$\Omega$. With the law held fixed the offsets lie between 0.61 and 0.65~k$\Omega$, constant to 3\%, while the inelastic amplitude grows by a factor of 2.4--2.7 over the same range. With amplitude and offset both left free per curve (Fig.~\ref{fig:slopes}b), the offset stays within 5\% from 0.20 to 0.30~V/nm and falls by 8 and 22\% at 0.34 and 0.38~V/nm, where the low-temperature traces are least reliable, while the amplitude grows 2.6-fold; one shared offset with free amplitudes fits the five curves at 9.8~$\Omega$ rms against 4.6~$\Omega$ with free offsets. What the offset says about the carriers depends on the elastic scatterers. In $R=\mathcal K(\Gamma_c+\Gamma_{\rm in})$ with $\mathcal K\propto1/E_F$, charged impurities give $R_{\rm el}\propto1/n_d$, and a constant offset then holds $n_d$ to a few percent, while short-range scatterers give $\Gamma_c\propto E_F$ and an offset independent of $n_d$. The inelastic part tests the number in either case, because the borrowed rate does not depend on $n_d$ and $R_{\rm in}\propto\mathcal K(D)\langle|V|^2\rangle$.

Below 0.20~V/nm a second feature appears. Extrapolating the law to $D=0.14$ and 0.10~V/nm and fitting only an offset and an amplitude factor, the amplitude factors come out as 1.01 and 1.11 when the law is fixed by all five high-field curves, and between 0.95 and 1.21 when either end curve is left out, so the inelastic part is predicted to within about 20\%. The offsets, however, are 0.47 and 0.36~k$\Omega$, 25--45\% below the high-field value. The raw resistance at 5~K drops in the same way, from 0.73--0.86~k$\Omega$ at 0.20--0.30~V/nm to 0.58~k$\Omega$ at 0.14, although superconducting fluctuations may contribute there. In the Matthiessen form $R=\mathcal K(D)[\Gamma_c(D)+\Gamma_{\rm in}(T,D)]$ a change of the Dirac carrier number or mass enters through $\mathcal K$, and for charged scatterers through $\Gamma_c$ as well. Between 0.10 and 0.38~V/nm the inelastic part follows $D^2+D_0^2$ to 5--20\% with $\mathcal K$ fixed, depending on which high-field curves fix the law, while the elastic part changes by about 40\%. The two parts move independently, and together with the bound on the quartic term this limits any change of $\mathcal K$ to 10--20\% over the whole data set. Equation~(\ref{eq:V}) does not produce the rise of the elastic rate between 0.10 and 0.20~V/nm. It is also where the offset is most degenerate with the background of Sec.~\ref{sec:background} and where superconducting fluctuations may enter the low-temperature resistance, so we use it only as evidence that the two parts move separately.

Self-doping~\cite{LiangZhaoZhang2026} shifts the Dirac cone by an amount $\propto D^2$ at small field, so its leading field dependence is also quadratic, and the $D^2$ law alone cannot separate it from admixture. What separates them is what each changes. A shift of the Dirac carrier number changes $\mathcal K$, which the shape of the slope law holds to about 10\%. Carriers doped into the flat band change its entropy by $(\partial S/\partial\nu)\,\delta\nu$, which is large near the fillings where the Pomeranchuk effect is seen~\cite{Rozen2021,Saito2021}, whereas admixture leaves it fixed to order $N_d/N_f$. A self-doping law in which the Dirac pocket empties at a field $D_c$, of the form $a/(1-D^2/D_c^2)^2$, gives $\chi^2=168$, and the quartic bound requires $D_c\gtrsim1.2$~V/nm, a change of $\mathcal K$ below about 10\% at 0.38~V/nm. Only a $\mathcal K$ that itself grows as $1+D^2/D_0^2$, with $1/E_F$ rising fivefold between 0.10 and 0.38~V/nm, reproduces the slope law at a constant elastic part, and the Dirac oscillation frequency of Table~\ref{tab:tests} tests it directly.

\subsection{The thermal average and $D_0$}\label{sec:background}

Fitting the exact Eq.~(\ref{eq:rho}) to the five curves at $D\geq0.20$~V/nm, with a shared elastic rate, $\langle|V|^2\rangle\propto D^2+D_0^2$ and shared $\kappa$ and $\Delta$, uses five parameters for five curves. A single flat level reaches 12.5--12.9~$\Omega$ rms, but only by sending $D_0\to0$ and building a $-19$ to $-22\%$ quartic term out of the thermal average, and extrapolated to the two lowest fields it then misses their inelastic parts by factors of 1.3--2.0. Symmetric weight at $\pm\Delta$ gives 16.4--16.7~$\Omega$, a quartic share below 0.3\%, $D_0=0.155\pm0.004$~V/nm, and inelastic parts at the two lowest fields within 10--15\%. Without a real part of $\Sigma_f$ it gives $\kappa=3.2\pm0.2$ and $\Delta=225\pm13$~K (19~meV); with a marginal real part the fit runs to the limit of frequency-independent $g$, which is Eq.~(\ref{eq:F}). The low-field curves therefore disfavor a single flat level and, of the two level structures tested, select the particle--hole-symmetric weight expected near $\nu=2$, with a flat-band width, $\kappa\gtrsim3$, large enough that the zero-frequency profile reproduces the shape and the field-independent maximum. The maximum itself lies below $\Delta/\kappa$ at this $\kappa$, at 56.6~K against 71~K for the symmetric fit, so $\Delta$ is read from the exact fit rather than from $\kappa k_BT^*$.

The two lowest-field curves keep rising above $T^*$ (Fig.~\ref{fig:collapse}b). A field-independent monotonic contribution, from phonon scattering of the Dirac carriers or a parallel channel, does this, because its relative weight is largest where $W$ is smallest. A shared linear term $b\,T$ in the eight-curve fit with free offsets lowers the rms from 14.2 to 10.2~$\Omega$, and the residuals of the two lowest curves from 23 and 17 to 11 and 8~$\Omega$, with $b=1.8$~$\Omega$/K, $T^*=52.6$~K and $D_0=0.104$~V/nm. Restricted to $D\geq0.20$~V/nm the term is small, $b=0.27\pm0.04$~$\Omega$/K, and $D_0=0.136$~V/nm. A field-independent background competes with the $D_0^2$ term, since both are field-independent parts of the temperature dependence. The treatments above, slopes, profile amplitudes with and without background and the exact average, span $D_0=0.10$--$0.18$~V/nm (slopes 0.129, profile with background 0.104--0.136, exact average 0.155, four-parameter law 0.184), and we quote $D_0=0.15\pm0.03$~V/nm. The total zero-field slope is not affected by this degeneracy, because the slopes at 0.10 and 0.14~V/nm are measured: $D_0=0.10$ or 0.18~V/nm with the slope at 0.30~V/nm held fixed would miss the 0.14~V/nm slope by 2.7 and 4.7 standard deviations. The share of the zero-field slope carried by the admixture lies between about 1.8 and 3.4~$\Omega$/K, and a curve at $D=0$ would fix it.

If the mirror-breaking disorder were smooth on the scale of $\ell$, the sample would be a network of local resistivities $R_{\rm el}+W_2|D+h(\mathbf r)|^2F$ rather than a homogeneous medium. A two-dimensional Bruggeman effective medium with complex Gaussian $h$ lowers the 12--32~K slopes by 13--18\%, which the fitted amplitude absorbs, leaves the apparent $D_0$ unchanged to 2\%, and adds a quartic share of $+8\%$ at 0.38~V/nm. The measured quartic share, $-6\pm10\%$, is 1.4 standard deviations away, so the data slightly favor mirror breaking that is short-ranged on the scale of the mean free path.

A Kondo coupling $J\propto D^2$ between Dirac carriers and flat-band moments would move a coherence scale exponentially, while $T^*$ stays within 3~K from 0.20 to 0.38~V/nm as the amplitude grows 2.5-fold; its Born-level slope $\propto D^4$ also fits poorly (rms 1.9 against 0.3~$\Omega$/K). Because the Dirac density of states is small, any Kondo scale generated by the Dirac carriers is exponentially small, and the maximum is better read as a property of the flat band, seen by the Dirac carriers through Eq.~(\ref{eq:sigma}). A lattice coherence scale set by intersite exchange or by the flat-band width would not move exponentially, but it is then a flat-band scale, which is what Eq.~(\ref{eq:F}) describes. Phonon scattering with a $D$-dependent Dirac mass gives $a(1+\kappa D^2)^2$, with $\chi^2=60$, and no field-independent maximum. Phonons do give a large linear resistivity in magic-angle bilayers~\cite{Polshyn2019,WuHwangDasSarma2019}, while terahertz photoresistance points to electron--electron scattering in the linear regime near the magic angle~\cite{Shilov2026}. A van Hove crossing would put a peak in the slope as a function of $D$, as in twisted double bilayer graphene~\cite{Chu2022TDBG}; none appears up to 0.38~V/nm. A field-dependent parallel conductance would rescale the temperature-dependent part by $(R/R_s)^2$ along with the offset, which the low-field curves exclude in the same way as a change of $\mathcal K$.

Device D1 ($\theta=1.52^\circ$) appears only as colour maps of $R_{xx}(\nu,T)$ at $D=0$, 0.1, 0.2 and 0.3~V/nm (Supplemental Fig.~S4 of Ref.~\cite{Zhang2026PRL}). Its twist angle is close to the magic angle of the trilayer, about $1.5^\circ$~\cite{Khalaf2019}, and it is insulating-like at $\nu=2.1$ for $D\geq0.1$, where a field-tuned correlated state can change the flat band itself, so it lies outside the regime treated here. At $\nu=2.3$--2.5, windows starting at 12~K favour $p=2$ with $D_0=0.12$--$0.15$~V/nm, windows starting at 15--20~K favour $p=1$, and the maximum moves from about 70 to 50~K between $D=0.1$ and 0.3. A field-independent background moves the apparent maximum upward where the admixture is weak; in D3 the raw maximum moves from 71~K at 0.14~V/nm to 58~K at 0.38~V/nm and the background fit reproduces this, so a moving maximum is not by itself evidence against Eq.~(\ref{eq:factor}). The same Supplemental Material reports a Dirac cyclotron mass in D1 at $\nu=3$ that rises from 0.005 to 0.008\,$m_0$ between $D=0$ and 0.2~V/nm while the Dirac pocket shrinks. A shrinking cone of fixed velocity would lower the mass, so the velocity is renormalized, in the direction of Eq.~(\ref{eq:mass}). The rise of 60\% is larger than the 10--30\% that $1+Y$ gives, possibly because the D1 pocket at $\nu=3$ approaches depletion.

The Dirac pocket of D3 has not been reported, so we use that of D1 at $\nu=3$~\cite{Zhang2026PRL} ($F\approx1.1$~T, $m\approx0.005$--$0.007\,m_0$). With $R\approx\rho$, the inelastic Dirac rate at $D=0.38$~V/nm and $T=T^*$ is $\hbar\Gamma_d\approx2$--3~meV. Since $g(T^*)=1/2\Delta$, the weight is $Y=2\hbar\Gamma_d/\Delta$; for the symmetric fit, $\Delta\approx19$~meV, this gives $Y(0.38\,{\rm V/nm})\approx0.3$ and $Y(0)\approx0.04$, and for $\kappa=1$ it would be about 1. The rms hybridization is then about 10~meV at 0.38~V/nm, so $v_D\approx25$~meV per V/nm and $D_0$ corresponds to a local mirror-breaking energy $s\approx4$~meV; we use Eq.~(\ref{eq:V2}) as a parametrization, and $\pm D$ data separate a quenched $s$ from a uniform offset. The hybridization is thus a sizable fraction of $\Delta$ at the largest field, and the quadratic law survives because of the structure of Eq.~(\ref{eq:sigma}) rather than because $V$ is small. $Y$ sets the static mass and the $O(\Delta^{-4})$ terms of Eq.~(\ref{eq:lowE}), which the exact average of Sec.~\ref{sec:background} keeps, whereas the feedback parameter sets the corrections to Eq.~(\ref{eq:sigma}) itself. With $N_d\approx4.5$~eV$^{-1}$ per moir\'e cell the feedback parameter is 0.02--0.03 for $\kappa=1$--3.

\subsection{Own and borrowed dissipation}\label{sec:entropy}

A local critical bath enters the electrons' self-energy through an Eliashberg function $\alpha^2F(\Omega,T)=\frac{\lambda}{2}w(\Omega/k_BT)$ with $w\to1$ above $k_BT$, and its own electrons then scatter at the rate $-{\rm Im}\,\Sigma=\frac{\pi}{2}\lambda k_BT\,\phi(\varepsilon/k_BT)$, with
\begin{equation}
\phi(x)=\int_0^\infty\! dy\,w(y)\Big[\coth\frac y2-\frac12\tanh\frac{y+x}{2}-\frac12\tanh\frac{y-x}{2}\Big]
\label{eq:phi}
\end{equation}
and $\phi\to|x|$ at large $|x|$~\cite{SkrgicStojkovic2026}. The bath is summarized by its exponent, its transport prefactor and its entropy coefficient. The exponent $\beta=\phi(0)=2\int_0^\infty dy\,w/\sinh y$ grows as the bath relaxes more slowly. The Kubo average of the same rate gives $\alpha=C_b\lambda$ with $C_b=\pi/\langle1/\phi\rangle$, the average taken with $-\partial n_F/\partial x=1/4\cosh^2(x/2)$, and $C_b\simeq3.6\,\beta$ to within 5\% across six families of shapes~\cite{SkrgicStojkovic2026}. The logarithm in $C/T$ is $B=b\,\gamma_b\lambda$ with
\begin{equation}
b=1+\frac{3I_w}{\pi^2},\qquad I_w=\int_0^\infty dy\,y\,[1-w(y)],
\label{eq:bown}
\end{equation}
the electrons' own mass logarithm plus the entropy of the bath's thermal response. The ratio $Q=\alpha\gamma_b/B=C_b/b=C_b\,\eta$ is free of $\lambda$, and $\eta=1/b$ is the share of the critical entropy carried by current-relaxing scattering. A conformal bath, $w=\tanh(y/2)$, has $\beta=2$, $b=3/2$ and $Q=4\pi^3/21\zeta(3)\approx4.91$. These own-channel results are those of Ref.~\cite{SkrgicStojkovic2026}; the borrowed channel below is new.

\begin{figure}[t]
\includegraphics[width=0.95\columnwidth]{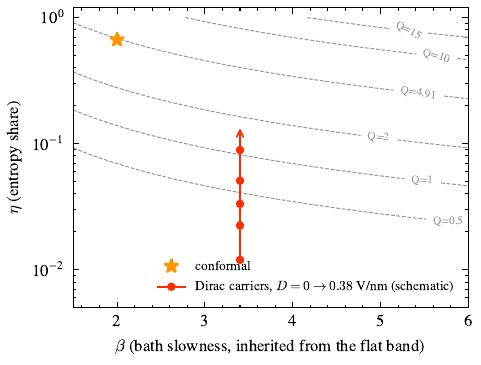}
\caption{The $(\beta,\eta)$ map with contours of $Q\simeq3.6\,\beta\eta$ (dashed) and the conformal point (star). The Dirac carriers inherit the flat band's $\beta$ and their entropy share grows as $D^2+D_0^2$, so $D$ moves them vertically. The flat band's $\beta$ is not known and is drawn at an arbitrary value; the vertical positions use $\eta_d(0)=0.012$ and $D_0=0.15$~V/nm for illustration.}
\label{fig:map}
\end{figure}

By Eq.~(\ref{eq:lowE}), ${\rm Im}\,\Sigma_d=Y\,{\rm Im}\,\Sigma_f$, so the Dirac carriers see the flat band's spectral shape at every field, $\beta_d=\beta_f$, with coupling $\lambda_d=Y\lambda_f$. The logarithm in the total $C/T$ is dominated by the flat band, since $N_f\gg N_d$, and does not depend on $D$ at leading order. The Dirac carriers therefore carry an entropy share
\begin{equation}
\eta_d=\frac{\gamma_{b,d}\,Y\lambda_f}{B}\propto D^2+D_0^2,\qquad Q_d=C_b(\beta_f)\,\eta_d,
\label{eq:eta}
\end{equation}
with $\gamma_{b,d}$ the Dirac Sommerfeld coefficient, and the field moves them vertically on the $(\beta,\eta)$ map at fixed $\beta$ (Fig.~\ref{fig:map}). With $\gamma_{b,d}/\gamma_{b,f}\sim N_d/N_f\sim10^{-2}$ they sit far below the conformal point, $\eta_d\sim10^{-2}$--$10^{-1}$, and between $D=0$ and 0.38~V/nm the field moves them by $(D^2+D_0^2)/D_0^2\approx7$. Doping or pressure moves a cuprate or a heavy-fermion compound on this map only by changing its bath as well.

Differences between fields remove the band factor and the entropy of the flat band at once. Define
\begin{equation}
Q_{\rm dial}\equiv\gamma_{b,d}\,\frac{\partial\alpha_d/\partial(D^2)}{\partial B/\partial(D^2)} ,
\label{eq:Qdial}
\end{equation}
with $\alpha_d$ the slope of the Dirac transport rate, $\hbar/\tau_{\rm tr}=\alpha_dk_BT$. Phonons, the Dirac carriers' own scattering, the flat-band entropy and the field-independent band structure all cancel, and so does the coupling. The Dirac band factors reduce to a single number: with $\mathcal K$ of Appendix~\ref{app:kubo} and $\gamma_{b,d}=\frac{\pi^2}{3}k_B^2N_d$, the numerator is $\gamma_{b,d}\alpha_d=(\pi ge^2k_B/6\hbar)\,n_d\,\partial\rho/\partial T$, in which the Dirac velocity and Fermi energy cancel and only the Dirac density $n_d$, one quantum-oscillation frequency, remains. The denominator can be computed exactly at order $\langle|V|^2\rangle$, because the field-induced free energy is a single Dirac line closed by the full flat-band propagator (Appendix~\ref{app:bound}),
\begin{equation}
\begin{split}
\Delta\Omega=-N_d\langle|V|^2\rangle\!\int\! d\varepsilon\,A_f(\varepsilon,T)\Big[&\ln\frac{\Lambda}{2\pi k_BT}\\
&-{\rm Re}\,\psi\Big(\frac12+\frac{i\varepsilon}{2\pi k_BT}\Big)\Big],
\end{split}
\label{eq:Omega}
\end{equation}
with $A_f=-{\rm Im}\,G_f/\pi$ at the Dirac momentum and $\Lambda$ the Dirac bandwidth. For symmetric weight $A_f=|{\rm Im}\,\Sigma_f|/\pi\Delta^2$ at low energy, so $A_f$ is proportional to the same $\phi$ that sets the flat band's own rate. The $|x|$ part of $\phi$ reproduces the logarithm of the Dirac mass in Eq.~(\ref{eq:mass}); the thermal excess $\phi-|x|$ adds a second logarithm, carried by the flat band's response to heating. Together they give
\begin{equation}
b_{\rm dial}=1+\frac{3I_0}{\pi^2},\qquad I_0=\int_{-\infty}^{\infty}dx\,\big[\phi(x)-|x|-c'\big],
\label{eq:bdial}
\end{equation}
for the field-induced logarithm in units of the Dirac mass logarithm, $\partial B/\partial D^2=b_{\rm dial}\,\gamma_{b,d}\,\partial\lambda_d/\partial D^2$. Here $c'=\lim_{x\to\infty}(\phi-x)$ is a temperature-linear floor in the flat-band rate far from the Fermi level; it adds no logarithm, and it vanishes for the conformal bath. The Kubo average of the borrowed rate is the same functional as for the flat band's own electrons, so $\alpha_d=C_b\lambda_d$ with the same $C_b$, and
\begin{equation}
Q_{\rm dial}=\frac{C_b}{b_{\rm dial}} ,
\label{eq:Qdial2}
\end{equation}
independent of $\lambda_f$, $Y$, $\Delta$ and $v_D$. The full numerical evaluation of Eq.~(\ref{eq:Omega}) agrees with Eq.~(\ref{eq:bdial}) to 0.1\% and does not depend on the cutoff or the temperature window.

At second order in the hybridization the two entropy coefficients are tied exactly. The slope of $\phi$ is the bath spectrum smoothed by the Fermi window, and an integration by parts (Appendix~\ref{app:bound}) splits $I_0$ into the Fermi function's own thermal excess, $\pi^2/3$, and twice the bath's, $2I_w$. Hence
\begin{equation}
b_{\rm dial}=2\,b,\qquad Q_{\rm dial}=\frac{Q}{2},\qquad \eta_{\rm dial}=\frac{\eta}{2},
\label{eq:two}
\end{equation}
for every bath with dynamics of its own, that is, every $w(y)$ fixed by the bath and not by the electrons, with a convergent $I_w$. The identity needs neither a monotonic spectrum nor a vanishing floor. Borrowed dissipation thus keeps the exponent and the transport prefactor of the bath's own electrons and carries exactly twice their entropy per unit of coupling, although only a small fraction of the flat band's total entropy: the Dirac carriers reach the bath only through a flat-band fermion, and the thermal smearing of that fermion enters the entropy in full. Since $b\ge1$ for any $w\le1$, the borrowed share never exceeds $\eta_{\rm dial}=1/2$.

A flat band whose electrons scatter off a marginal local bath with spectrum $\chi''\propto\tanh(\omega/2k_BT)$~\cite{Varma1989,PatelMcGreevyArovasSachdev2018} has $\phi=x\coth(x/2)$ exactly. A flat spectrum of this kind appears, above the spin-glass scale, in random Heisenberg magnets near Sachdev--Ye--Kitaev criticality~\cite{Hosseinabadi2026}. Its thermal excess integrates to $I_0=2\pi^2/3$, so $b_{\rm dial}=3$, twice the $3/2$ of its own electrons, and $C_b=2\pi^3/7\zeta(3)$. Then
\begin{equation}
Q_{\rm dial}^{\rm conf}=\frac{2\pi^3}{21\zeta(3)}\approx2.46 .
\label{eq:Qconf}
\end{equation}

The identity fails for a bath that is a field damped by the electrons it scatters, the Yukawa--SYK class~\cite{Esterlis2021,PatelLuntsSachdev2024,PatelGuoEsterlisSachdev2023}. There the critical field acquires a thermal mass $m\propto T$, and its spectrum approaches the plateau only as $1-w\approx2m/\pi\Omega$, so $I_w$ grows as the bandwidth over $T$ instead of converging. For the field's own electrons this growth is cancelled by the field's free energy, and $b=1+\delta$ with $\delta=0.007$--$0.074$~\cite{SkrgicStojkovic2026}. The borrowing carriers have no such partner: the thermal mass lowers the flat-band rate at every energy in the band, and Eq.~(\ref{eq:Omega}) picks this up in full. On the large-$M$ saddle of the two-dimensional field, the field-induced $C/T$ rises as a power of $1/T$ over $k_BT/E_L=10^{-5}$--$10^{-2}$ instead of logarithmically, and its size and even its sign depend on how far the flat band extends. There is then no $b_{\rm dial}$. The transport side is untouched, since $C_b$ involves only $|\varepsilon|\sim k_BT$, so the distinction lies entirely in the entropy: a straight logarithm in $\partial(C/T)/\partial D^2$ marks a bath with dynamics of its own, and a curved, non-logarithmic one marks a self-damped field.

\begin{figure}[t]
\includegraphics[width=\columnwidth]{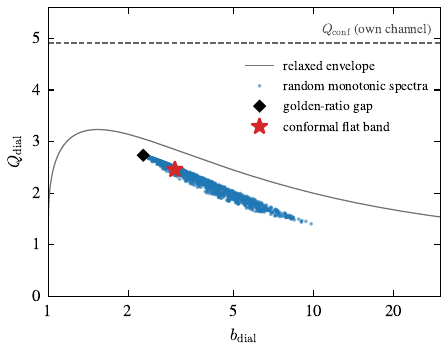}
\caption{Field-differential ratio $Q_{\rm dial}$ against $b_{\rm dial}$, Eqs.~(\ref{eq:bdial}) and (\ref{eq:Qdial2}). Dots: 1500 random monotonic bath spectra with no high-energy floor, built as mixtures of thermal gaps; they fill a narrow band that begins at the diamond. Diamond: their maximum, $Q_{\rm dial}=2.73$ at $b_{\rm dial}=2.28$, a single gap at $2\ln[(1+\sqrt5)/2]\,k_BT$. Star: the conformal flat band, $\phi=x\coth(x/2)$. Thin line: the relaxed envelope obtained by keeping only $\phi\ge|x|$, reached by $\phi^*=\max(|x|,k/2\cosh\frac{x}{2})$, which no monotonic spectrum produces; its maximum is 3.23 at $b_{\rm dial}=1.54$. Dashed: the conformal value $4.91$ in the bath's own channel. By Eq.~(\ref{eq:two}) every point has exactly twice its $Q_{\rm dial}$ in the own channel.}
\label{fig:bound}
\end{figure}

Other flat bands give other numbers, and a ceiling follows from two properties of $\phi$. For any bath spectrum that grows monotonically with frequency, $\phi$ is convex and lies above its asymptotes, $\phi\ge|x|+c'$, with $c'=\int_0^\infty dy\,[w(y)\coth(y/2)-1]$. A bath that relaxes faster than the conformal one, with less weight below $k_BT$ and a smaller $\beta$, has $c'<0$; a slower bath has $c'>0$. Where $c'=0$ the thermal broadening is confined to energies of order $k_BT$. Every non-decreasing spectrum is a superposition of steps with non-negative weights, and $\phi$ is linear in those weights, so the maximum of Eq.~(\ref{eq:Qdial2}) over realizable flat bands is a search over step mixtures with zero total floor (Appendix~\ref{app:bound}). A numerical search finds it at a single step, the one gap whose floor vanishes, at $s_0=2\ln[(1+\sqrt5)/2]\approx0.96$, and gives
\begin{equation}
Q_{\rm dial}\le2.73 ,
\label{eq:ceiling}
\end{equation}
with $b_{\rm dial}=2.28$ and $C_b=6.24$ at the maximum (Fig.~\ref{fig:bound}). The conformal flat band reaches 90\% of this ceiling. Keeping only $\phi\ge|x|$ and dropping the convexity that a monotonic spectrum imposes relaxes the bound to 3.23, reached by a shape that no monotonic spectrum produces. A negative floor only lowers the ceiling, so it holds for every faster bath as well (Appendix~\ref{app:bound}). A positive floor raises the relaxed ceiling, to 4.5 for $c'=0.5$ and 6.2 for $c'=1$, because it adds current relaxation at high energy without adding a logarithm; such a floor would itself show up as a temperature-linear flat-band rate far from the Fermi level.

Through Eq.~(\ref{eq:two}) the ceiling carries over to a bath acting on its own electrons: for monotonic spectra with $c'\le0$, $Q\le5.47$, reached by the same gap with $b=1.14$. This sharpens the conformal value 4.91, which is a single point of that class. Slower baths with $c'>0$ lie outside it; $w=\tanh y$, $1-e^{-y}$ and $\min(1,y)$ give $Q=10.0$, 7.7 and 11.7 in their own channel. A measured $Q$ above 5.47 therefore calls for a bath slower than conformal, a non-monotonic spectrum or a self-damped field.

Any further field-induced entropy that does not relax the Dirac current, for example from a field-dependent Dirac pocket, only adds to the denominator of Eq.~(\ref{eq:Qdial}). In the admixture picture $Q_{\rm dial}$ is therefore at most 2.46 for a conformal flat band and, by the numerical search, at most 2.73 for any monotonic flat-band bath without a positive high-energy floor. A value above 2.73 would mean that the flat-band bath is slower than conformal with a positive high-energy floor, which raises the relaxed ceiling to 4.5--6.2 for $c'=0.5$--1, that the field removes entropy elsewhere, as self-doping out of a Pomeranchuk-enhanced flat band could, or that the field-induced scattering is not hybridization. Under self-doping $\partial B/\partial D^2$ is set by $(\partial S/\partial\nu)(\partial\nu_f/\partial D^2)$, which has no reason to track $\partial\alpha_d/\partial D^2$ at all. Equations~(\ref{eq:lowE}), (\ref{eq:mass}) and (\ref{eq:Omega})--(\ref{eq:ceiling}) assume $|\Sigma_f|\ll\Delta$, that is $T\ll T^*$, whereas the transport analysis of Sec.~\ref{sec:data} uses the full $G_f$ and is not limited in this way. The ratio is therefore measured below about $T^*/5\approx10$~K and above the superconducting transition at this filling, and its numerator has to be measured in the same window; the 12--32~K slopes analyzed here do not supply it. The transport curves fix $\Delta$ and $\kappa$ but not the bath class, $\beta$, $c'$ or a marginal ${\rm Re}\,\Sigma_f$; that is what the entropy measures.

\section{Discussion and outlook}\label{sec:tests}

The transport side of the dial has been measured once, because displacement fields in MATTG have been swept to map superconductivity, and the normal-state curves that separate rate from number appear only as a supplemental figure. To our knowledge, the entropy side has not been measured at all. Entropy in moir\'e graphene has been obtained from ${\rm d}\mu/{\rm d}T$ as a function of filling and in-plane field~\cite{Rozen2021,Saito2021}, and from planar tunneling in the bilayer~\cite{ZhangAndrei2026}, but not at fixed filling in steps of $D$, which is the measurement that turns the classification into a controlled experiment. We present the decisive measurements in Table~\ref{tab:tests}, alongside an outcome that would confirm the admixture reading and the outcome that would move the classification.

\begin{table*}[t]
\caption{Measurements that test the idea of borrowed dissipation in MATTG and classify its bath. Numbers refer to device D3 at $\nu=2.1$ unless stated.}
\label{tab:tests}
\begin{ruledtabular}
\begin{tabular}{p{0.24\textwidth}p{0.36\textwidth}p{0.34\textwidth}}
measurement & admixture with a bath of its own & outcome that moves the reading\\
\hline
$R_{xx}(T)$ at $\pm D$ and at $D=0$, same device & even in $D$, no kink at 0; 12--32~K slope $3.4\pm0.3$~$\Omega$/K at $D=0$ and $22\pm1$~$\Omega$/K at $\pm0.30$~V/nm & minimum shifted to $D_{\rm off}$ (now between $-0.26$ and $+0.04$~V/nm): built-in mirror breaking; kink at 0: no hybridization\\
difference of two curves at 80--230~K & $x/(1+x^2)$ with the same $T^*$, free of any field-independent background & a different shape: the profile is not a flat-band property\\
Dirac oscillation frequency at $\nu=2.1$ versus $D$ & changes by at most about 10\% up to 0.38~V/nm; fixes $n_d$, the only band input of $Q_{\rm dial}$ & $1/E_F$ growing as $1+D^2/D_0^2$: self-doping\\
Dirac cyclotron mass versus $D$ and $T$ & static part grows by $1+Y\approx1.2$--1.3; logarithm with coefficient $\propto D^2+D_0^2$, Eq.~(\ref{eq:mass}) & no logarithm: quasi-elastic flat-band width\\
flat-band entropy at fixed $\nu$, from ${\rm d}\mu/{\rm d}T$ or $C/T$ in steps of $D$, $T\lesssim10$~K & field-independent to order $N_d/N_f$; field-induced part a straight logarithm in $T$ with $Q_{\rm dial}\le2.73$ for $c'\le0$ (conformal 2.46) & non-logarithmic: self-damped field; $Q_{\rm dial}>2.73$: slow bath with $c'>0$, self-doping or non-hybridization scattering\\
bilayer at $\theta_{\rm TTG}/\sqrt2$ and the same flat-band filling & own-channel $Q=2Q_{\rm dial}$ for the same bath, if the bilayer's flat-band electrons conduct as a band scattered by that bath & $Q\neq2Q_{\rm dial}$: the bath is self-damped, or the trilayer's Dirac cone screens the flat band\\
$D_0$ across devices & tracks the measured twist mismatch; for Gaussian $h$ the local $|V|$ is Rician, and decoupled points are rare as $e^{-D^2/D_0^2}$ & $D_0$ device-independent: intrinsic mirror breaking\\
filling away from $\nu=2$ & quartic term grows with $|\nu-2|$; maximum moves down as flat-band weight nears $E_F$ & no change: weight is not split into Hubbard bands\\
\end{tabular}
\end{ruledtabular}
\end{table*}

The first four tests use devices and gates that already exist. The ${\rm d}R/{\rm d}T$ slopes at $\pm D$ test the mirror structure of Eq.~(\ref{eq:V}) directly, and the superconducting map of Supplemental Fig.~S15a of Ref.~\cite{Zhang2026PRL} already spans both signs of $D$. The Dirac oscillation frequency is the decisive transport test: it separates admixture from self-doping by itself and supplies the one number the numerator of $Q_{\rm dial}$ needs. The cyclotron mass checks that the borrowed rate comes with its Kramers--Kronig partner, the logarithm in Eq.~(\ref{eq:mass}).

The entropy test is the one that places the bath. With $Y(0.38)\approx0.3$ and $\gamma_{b,d}/\gamma_{b,f}\sim10^{-2}$, the field-induced logarithm is of order a percent of the flat band's own. Reaching it with the ${\rm d}\mu/{\rm d}T$ technique of Refs.~\cite{Rozen2021,Saito2021} requires differencing in $D$ at fixed $\nu$, integrating $(\partial S/\partial n)_T$ from a reference filling, and separating the field-induced logarithm from changes of the flat band at order $D^2$ through remote bands and ordered states, which we have not estimated. A straight logarithm gives $b_{\rm dial}$ directly, and with the numerator gives $Q_{\rm dial}$; a curved one identifies a self-damped field without any value of $Q$.

The matched bilayer tests the identity between the two channels. The mirror-even sector of a trilayer at angle $\theta$ is a bilayer at $\theta/\sqrt2$ with interlayer coupling scaled by $\sqrt2$, so device D3 corresponds to a 0.90$^\circ$ bilayer, and a magic-angle bilayer to a trilayer near $1.5^\circ$. At the same flat-band filling, $\nu_f=\nu-\nu_d$, the continuum Hamiltonians agree up to an overall factor $\sqrt2$ in energy, while the trilayer's flat band is in addition screened by its Dirac cone. For a bath with dynamics of its own the bilayer's electrons should then show exactly twice the trilayer's $Q_{\rm dial}$. The bilayer's own logarithm is not yet available: near $\nu=\pm2$ its entropy is dominated by a local-moment contribution~\cite{Rozen2021,Saito2021} that is two orders of magnitude larger than the bath logarithm expected from its slope~\cite{Cao2020PRL,Jaoui2022}, and separating the two needs sub-kelvin entropy with the field-sensitive part subtracted.

The main lesson of this work is that a Planckian rate can be inherited. In a nutshell, a carrier that is not itself critical, hybridized with a sector that is, acquires that sector's temperature-linear rate with its spectral shape and its transport prefactor intact, and in MATTG the share it acquires is set by a gate, in analogy with the proximity effect~\cite{Zutic2019}. That conduction electrons inherit the flat-band self-energy is the structure of the periodic Anderson model, and dynamical mean-field studies of the heavy-fermion description of magic-angle graphene find it~\cite{Rai2024,Calderon2025}; what the displacement field adds is a knob on that inheritance, and the field-differential entropy is what makes it a measurement of the bath. The analyzed curves are consistent with this inheritance: the field changes how fast the Dirac carriers relax and, within the bound set by the slope law, leaves their number alone, with one temperature profile and a fixed maximum across a fourfold change in amplitude. A temperature-linear resistivity therefore does not by itself say which carriers are critical. The entropy does. Per unit of coupling, borrowed dissipation carries exactly twice the entropy of the bath's own electrons when the bath has dynamics of its own, and a non-logarithmic entropy when the bath is a field damped by the electrons it scatters. These two kinds of bath are the main alternatives in current theories of strange metals, and the trilayer can separate them without the coupling, through differences between fields in one device, provided the field-induced entropy is resolved between the superconducting transition and about 10~K.

The same question arises wherever light and heavy carriers coexist. Heavy-fermion phenomenology describes them as two fluids whose mixing is set by temperature~\cite{NakatsujiPinesFisk2004,Yang2016}; in MATTG light and heavy carriers are both visible in transport and thermodynamics~\cite{Pierce2025,Merino2025,Calugaru2024}, and their mixing is set by a gate. Van der Waals heterostructures are routinely used to lend one material the spin--orbit coupling, magnetism or pairing of its neighbor~\cite{Zutic2019}; here the neighbor is a strongly correlated flat band and what it lends is its dissipation. Flat bands can also be made in single graphene sheets, by doping toward a van Hove point or by patterning vacancies, where interactions turn a nodal-line semimetal into a magnetic insulator~\cite{Yu2026Flat,Liu2022Vacancy,MedeirosSilva2025}, and a light carrier placed beside such a band would borrow from it in the same way.

The strange-metal phenomenology of magic-angle graphene~\cite{Cao2020PRL,Polshyn2019,Jaoui2022,ZhangLi2026,HartnollMackenzie2022,Legros2019} here enters only through the Planckian width of the flat-band states, whatever its origin may be. The displacement field then reads that width out through a second, light species whose coupling can be turned, and on the bath diagram it moves a carrier vertically while holding its bath fixed. The field-differential ratio then measures the flat band's own bath, and the shape of the field-induced entropy says what kind of bath it is.

\begin{acknowledgments}
D.~\v{S}. and D.~S. are partially supported by NSF, under Grant No.~PHY-2310363. Claude Opus 5.5 was used in assisting with literature search, numerical analysis (including digitization of measurement plots), as well as text editing. The authors take full responsibility for the claims.
\end{acknowledgments}

\bibliography{refs_v7}

@misc{SkrgicMATTGcode,
  author       = {{\v{S}}krgi{\'c}, Dino and Stojkovic, Dejan},
  title        = {Code and digitized data for ``Borrowed {Planckian} dissipation in magic-angle twisted graphene''},
  year         = {2026},
  howpublished = {\url{https://github.com/dinoskrgic-ctrl/mattg}},
  url          = {https://github.com/dinoskrgic-ctrl/mattg},
  note         = {GitHub repository}
}

@article{Zhang2026PRL,
  author  = {Zhang, Le and Zhou, Wenqiang and Fang, Xinjie and Zhan, Zhen and others},
  title   = {Heavy-Fermion Phase Diagram in Magic-Angle Twisted Trilayer Graphene},
  journal = {Phys. Rev. Lett.}, volume = {137}, pages = {046301}, year = {2026},
  doi     = {10.1103/zzks-vkl2}, note = {arXiv:2507.12254}
}

@article{Khalaf2019,
  author  = {Khalaf, Eslam and Kruchkov, Alex J. and Tarnopolsky, Grigory and Vishwanath, Ashvin},
  title   = {Magic angle hierarchy in twisted graphene multilayers},
  journal = {Phys. Rev. B}, volume = {100}, pages = {085109}, year = {2019}, doi = {10.1103/PhysRevB.100.085109}
}

@article{Carr2020,
  author  = {Carr, Stephen and Li, Chenyuan and Zhu, Ziyan and Kaxiras, Efthimios and others},
  title   = {Ultraheavy and Ultrarelativistic {D}irac Quasiparticles in Sandwiched Graphenes},
  journal = {Nano Lett.}, volume = {20}, pages = {3030}, year = {2020}, doi = {10.1021/acs.nanolett.9b04979}
}

@article{Lei2021,
  author  = {Lei, Chao and Linhart, Lukas and Qin, Wei and Libisch, Florian and others},
  title   = {Mirror symmetry breaking and lateral stacking shifts in twisted trilayer graphene},
  journal = {Phys. Rev. B}, volume = {104}, pages = {035139}, year = {2021}, doi = {10.1103/PhysRevB.104.035139}
}

@article{Phong2021,
  author  = {Phong, V{\~o} Ti{\'{\^e}}n and Pantale{\'o}n, Pierre A. and Cea, Tommaso and Guinea, Francisco},
  title   = {Band structure and superconductivity in twisted trilayer graphene},
  journal = {Phys. Rev. B}, volume = {104}, pages = {L121116}, year = {2021}, doi = {10.1103/PhysRevB.104.L121116}
}

@article{Christos2022,
  author  = {Christos, Maine and Sachdev, Subir and Scheurer, Mathias S.},
  title   = {Correlated Insulators, Semimetals, and Superconductivity in Twisted Trilayer Graphene},
  journal = {Phys. Rev. X}, volume = {12}, pages = {021018}, year = {2022}, doi = {10.1103/PhysRevX.12.021018}
}

@article{RamiresLado2021,
  author  = {Ramires, Aline and Lado, Jose L.},
  title   = {Emulating Heavy Fermions in Twisted Trilayer Graphene},
  journal = {Phys. Rev. Lett.}, volume = {127}, pages = {026401}, year = {2021}, doi = {10.1103/PhysRevLett.127.026401}
}

@article{SongBernevig2022,
  author  = {Song, Zhi-Da and Bernevig, B. Andrei},
  title   = {Magic-Angle Twisted Bilayer Graphene as a Topological Heavy Fermion Problem},
  journal = {Phys. Rev. Lett.}, volume = {129}, pages = {047601}, year = {2022}, doi = {10.1103/PhysRevLett.129.047601}
}

@article{Yu2023THF,
  author  = {Yu, Jiabin and Xie, Ming and Bernevig, B. Andrei and Das Sarma, Sankar},
  title   = {Magic-angle twisted symmetric trilayer graphene as a topological heavy-fermion problem},
  journal = {Phys. Rev. B}, volume = {108}, pages = {035129}, year = {2023}, doi = {10.1103/PhysRevB.108.035129}
}

@article{HuBernevigTsvelik2023,
  author  = {Hu, Haoyu and Bernevig, B. Andrei and Tsvelik, Alexei M.},
  title   = {Kondo Lattice Model of Magic-Angle Twisted-Bilayer Graphene: {H}und's Rule, Local-Moment Fluctuations, and Low-Energy Effective Theory},
  journal = {Phys. Rev. Lett.}, volume = {131}, pages = {026502}, year = {2023}, doi = {10.1103/PhysRevLett.131.026502}
}

@article{ChouDasSarma2023,
  author  = {Chou, Yang-Zhi and Das Sarma, Sankar},
  title   = {Kondo Lattice Model in Magic-Angle Twisted Bilayer Graphene},
  journal = {Phys. Rev. Lett.}, volume = {131}, pages = {026501}, year = {2023}, doi = {10.1103/PhysRevLett.131.026501}
}

@article{LauColeman2025,
  author  = {Lau, Liam L. H. and Coleman, Piers},
  title   = {Topological Mixed Valence Model for Twisted Bilayer Graphene},
  journal = {Phys. Rev. X}, volume = {15}, pages = {021028}, year = {2025}, doi = {10.1103/PhysRevX.15.021028}
}

@misc{LiangZhaoZhang2026,
  author  = {Liang, Bokai and Zhao, Jing-Yu and Zhang, Ya-Hui},
  title   = {Displacement-Field-Driven Semimetal-Superconductor Transition in Magic-Angle Twisted Trilayer Graphene},
  year    = {2026}, note = {arXiv:2606.20849}
}

@article{Park2021,
  author  = {Park, Jeong Min and Cao, Yuan and Watanabe, Kenji and Taniguchi, Takashi and Jarillo-Herrero, Pablo},
  title   = {Tunable strongly coupled superconductivity in magic-angle twisted trilayer graphene},
  journal = {Nature}, volume = {590}, pages = {249}, year = {2021}, doi = {10.1038/s41586-021-03192-0}
}

@article{Hao2021,
  author  = {Hao, Zeyu and Zimmerman, A. M. and Ledwith, Patrick and Khalaf, Eslam and others},
  title   = {Electric field--tunable superconductivity in alternating-twist magic-angle trilayer graphene},
  journal = {Science}, volume = {371}, pages = {1133}, year = {2021}, doi = {10.1126/science.abg0399}
}

@article{Kim2022,
  author  = {Kim, Hyunjin and Choi, Youngjoon and Lewandowski, Cyprian and Thomson, Alex and others},
  title   = {Evidence for unconventional superconductivity in twisted trilayer graphene},
  journal = {Nature}, volume = {606}, pages = {494}, year = {2022}, doi = {10.1038/s41586-022-04715-z}
}

@article{Zhou2025,
  author  = {Zhou, Zekang and Jiang, Jin and Karnatak, Paritosh and Wang, Ziwei and others},
  title   = {Double-dome unconventional superconductivity in twisted trilayer graphene},
  journal = {Nat. Phys.}, volume = {21}, pages = {1773}, year = {2025}, note = {arXiv:2404.09909}
}

@article{Shen2023,
  author  = {Shen, Cheng and Ledwith, Patrick J. and Watanabe, Kenji and Taniguchi, Takashi and others},
  title   = {Dirac spectroscopy of strongly correlated phases in twisted trilayer graphene},
  journal = {Nat. Mater.}, volume = {22}, pages = {316}, year = {2023}, doi = {10.1038/s41563-022-01428-6}
}

@article{Turkel2022,
  author  = {Turkel, Simon and Swann, Joshua and Zhu, Ziyan and Christos, Maine and others},
  title   = {Orderly disorder in magic-angle twisted trilayer graphene},
  journal = {Science}, volume = {376}, pages = {193}, year = {2022}, doi = {10.1126/science.abk1895}
}

@article{Pierce2025,
  author  = {Pierce, Andrew T. and Xie, Yonglong and Park, Jeong Min and Cai, Zhuozhen and others},
  title   = {Tunable interplay between light and heavy electrons in twisted trilayer graphene},
  journal = {Nat. Phys.}, volume = {21}, pages = {1237}, year = {2025}, doi = {10.1038/s41567-025-02956-z}
}

@article{Merino2025,
  author  = {Merino, Rafael Luque and C{\u{a}}lug{\u{a}}ru, Dumitru and Hu, Haoyu and D{\'i}ez-M{\'e}rida, Jaime and others},
  title   = {Interplay between light and heavy electron bands in magic-angle twisted bilayer graphene},
  journal = {Nat. Phys.}, volume = {21}, pages = {1078}, year = {2025}, doi = {10.1038/s41567-025-02912-x}
}

@misc{Calugaru2024,
  author  = {C{\u{a}}lug{\u{a}}ru, Dumitru and Hu, Haoyu and Merino, Rafael Luque and Regnault, Nicolas and Efetov, Dmitri K. and Bernevig, B. Andrei},
  title   = {The Thermoelectric Effect and Its Natural Heavy Fermion Explanation in Twisted Bilayer and Trilayer Graphene},
  year    = {2024}, note = {arXiv:2402.14057}
}

@article{Rozen2021,
  author  = {Rozen, Asaf and Park, Jeong Min and Zondiner, Uri and Cao, Yuan and others},
  title   = {Entropic evidence for a {P}omeranchuk effect in magic-angle graphene},
  journal = {Nature}, volume = {592}, pages = {214}, year = {2021}, doi = {10.1038/s41586-021-03319-3}
}

@article{Saito2021,
  author  = {Saito, Yu and Yang, Fangyuan and Ge, Jingyuan and Liu, Xiaoxue and others},
  title   = {Isospin {P}omeranchuk effect in twisted bilayer graphene},
  journal = {Nature}, volume = {592}, pages = {220}, year = {2021}, doi = {10.1038/s41586-021-03409-2}
}

@article{Cao2020PRL,
  author  = {Cao, Yuan and Chowdhury, Debanjan and Rodan-Legrain, Daniel and Rubies-Bigorda, Oriol and others},
  title   = {Strange Metal in Magic-Angle Graphene with near {P}lanckian Dissipation},
  journal = {Phys. Rev. Lett.}, volume = {124}, pages = {076801}, year = {2020}, doi = {10.1103/PhysRevLett.124.076801}
}

@article{Polshyn2019,
  author  = {Polshyn, Hryhoriy and Yankowitz, Matthew and Chen, Shaowen and Zhang, Yuxuan and others},
  title   = {Large linear-in-temperature resistivity in twisted bilayer graphene},
  journal = {Nat. Phys.}, volume = {15}, pages = {1011}, year = {2019}, doi = {10.1038/s41567-019-0596-3}
}

@article{WuHwangDasSarma2019,
  author  = {Wu, Fengcheng and Hwang, Euyheon and Das Sarma, Sankar},
  title   = {Phonon-induced giant linear-in-{$T$} resistivity in magic angle twisted bilayer graphene: Ordinary strangeness and exotic superconductivity},
  journal = {Phys. Rev. B}, volume = {99}, pages = {165112}, year = {2019}, doi = {10.1103/PhysRevB.99.165112}
}

@article{Jaoui2022,
  author  = {Jaoui, Alexandre and Das, Ipsita and Di Battista, Giorgio and D{\'i}ez-M{\'e}rida, Jaime and others},
  title   = {Quantum critical behaviour in magic-angle twisted bilayer graphene},
  journal = {Nat. Phys.}, volume = {18}, pages = {633}, year = {2022}, doi = {10.1038/s41567-022-01556-5}
}

@article{ZhangLi2026,
  author  = {Zhang, Naiyuan and Nosov, Pavel and Sommer, Ofer and Wang, Yibang and others},
  title   = {Angular interplay of nematicity, superconductivity and strange metallicity in magic-angle twisted trilayer graphene},
  journal = {Nat. Phys.}, volume = {22}, pages = {527}, year = {2026}, doi = {10.1038/s41567-026-03202-w}
}

@article{Shilov2026,
  author  = {Shilov, A. L. and Elesin, L. and Kravtsov, M. and Covey, J. and Ganichev, S. D. and Kashchenko, M. A. and others},
  title   = {Interaction-limited conductivity of twisted bilayer graphene revealed by giant terahertz photoresistance},
  journal = {Nat. Commun.}, volume = {17}, pages = {9734}, year = {2026}, doi = {10.1038/s41467-026-76554-9}
}

@article{Chu2022TDBG,
  author  = {Chu, Yanbang and Liu, Le and Shen, Cheng and Tian, Jinpeng and others},
  title   = {Temperature-linear resistivity in twisted double bilayer graphene},
  journal = {Phys. Rev. B}, volume = {106}, pages = {035107}, year = {2022}, doi = {10.1103/PhysRevB.106.035107}
}

@article{Uri2020,
  author  = {Uri, A. and Grover, S. and Cao, Y. and Crosse, J. A. and others},
  title   = {Mapping the twist-angle disorder and {L}andau levels in magic-angle graphene},
  journal = {Nature}, volume = {581}, pages = {47}, year = {2020}, doi = {10.1038/s41586-020-2255-3}
}

@article{Wilson2020,
  author  = {Wilson, Justin H. and Fu, Yixing and Das Sarma, S. and Pixley, J. H.},
  title   = {Disorder in twisted bilayer graphene},
  journal = {Phys. Rev. Research}, volume = {2}, pages = {023325}, year = {2020}, doi = {10.1103/PhysRevResearch.2.023325}
}

@article{Georges1996,
  author  = {Georges, Antoine and Kotliar, Gabriel and Krauth, Werner and Rozenberg, Marcelo J.},
  title   = {Dynamical mean-field theory of strongly correlated fermion systems and the limit of infinite dimensions},
  journal = {Rev. Mod. Phys.}, volume = {68}, pages = {13}, year = {1996}, doi = {10.1103/RevModPhys.68.13}
}

@article{PJF1995,
  author  = {Pruschke, Thomas and Jarrell, Mark and Freericks, James K.},
  title   = {Anomalous normal-state properties of high-{$T_c$} superconductors: intrinsic properties of strongly correlated electron systems?},
  journal = {Adv. Phys.}, volume = {44}, pages = {187}, year = {1995}, doi = {10.1080/00018739500101526}
}

@book{Hewson1993,
  author    = {Hewson, A. C.},
  title     = {The {K}ondo Problem to Heavy Fermions},
  publisher = {Cambridge University Press}, address = {Cambridge}, year = {1993}, doi = {10.1017/CBO9780511470752}
}

@article{Yang2016,
  author  = {Yang, Yi-feng},
  title   = {Two-fluid model for heavy electron physics},
  journal = {Rep. Prog. Phys.}, volume = {79}, pages = {074501}, year = {2016}, doi = {10.1088/0034-4885/79/7/074501}
}

@article{NakatsujiPinesFisk2004,
  author  = {Nakatsuji, S. and Pines, D. and Fisk, Z.},
  title   = {Two Fluid Description of the {K}ondo Lattice},
  journal = {Phys. Rev. Lett.}, volume = {92}, pages = {016401}, year = {2004}, doi = {10.1103/PhysRevLett.92.016401}
}

@article{HartnollMackenzie2022,
  author  = {Hartnoll, Sean A. and Mackenzie, Andrew P.},
  title   = {Colloquium: {P}lanckian dissipation in metals},
  journal = {Rev. Mod. Phys.}, volume = {94}, pages = {041002}, year = {2022}, doi = {10.1103/RevModPhys.94.041002}
}

@article{Legros2019,
  author  = {Legros, A. and Benhabib, S. and Tabis, W. and Lalibert{\'e}, F. and others},
  title   = {Universal {$T$}-linear resistivity and {P}lanckian dissipation in overdoped cuprates},
  journal = {Nat. Phys.}, volume = {15}, pages = {142}, year = {2019}, doi = {10.1038/s41567-018-0334-2}
}

@article{Esterlis2021,
  author  = {Esterlis, Ilya and Guo, Haoyu and Patel, Aavishkar A. and Sachdev, Subir},
  title   = {Large-{$N$} theory of critical {F}ermi surfaces},
  journal = {Phys. Rev. B}, volume = {103}, pages = {235129}, year = {2021}, doi = {10.1103/PhysRevB.103.235129}
}

@article{PatelLuntsSachdev2024,
  author  = {Patel, Aavishkar A. and Lunts, Peter and Sachdev, Subir},
  title   = {Localization of overdamped bosonic modes and transport in strange metals},
  journal = {Proc. Natl. Acad. Sci. U.S.A.}, volume = {121}, pages = {e2402052121}, year = {2024}, doi = {10.1073/pnas.2402052121}
}

@misc{SkrgicStojkovic2026,
  author  = {\v{S}krgi\'c, Dino and Stojkovic, Dejan},
  title   = {Planckian dissipation per unit critical entropy as a coupling-free probe of strange-metal baths},
  year    = {2026}, note = {arXiv:2610.35939}
}

@article{Zutic2019,
  author = {{\v{Z}}uti{\'c}, Igor and Matos-Abiague, Alex and Scharf, Benedikt and Dery, Hanan and Belashchenko, Kirill},
  title = {Proximitized materials},
  journal = {Mater. Today}, volume = {22}, pages = {85}, year = {2019},
  doi = {10.1016/j.mattod.2018.05.003}
}

@article{Liu2022Vacancy,
  author = {Liu, Fujun and Qu, Fanyao and {\v{Z}}uti{\'c}, Igor and Malard, Mariana},
  title = {Vacancy-engineered nodal-line semimetals},
  journal = {Sci. Rep.}, volume = {12}, pages = {14981}, year = {2022},
  doi = {10.1038/s41598-022-18519-8}
}

@article{MedeirosSilva2025,
  author = {Medeiros-Silva, Andressa R. and Malard, Mariana and Pereira, Rodrigo G. and Paiva, Thereza},
  title = {Electronic interactions in a vacancy-engineered honeycomb lattice: Transition from a nodal-line semimetal to a magnetic insulator},
  journal = {Phys. Rev. B}, volume = {111}, pages = {195152}, year = {2025},
  doi = {10.1103/PhysRevB.111.195152}
}

@article{Yu2026Flat,
  author = {Yu, Chuanyu and Zheng, Tianhang and Liu, Fujun and Honda, Diego Elias and Malard, Mariana},
  title = {Tunable flat bands and van {H}ove singularities in doped graphene: A pathway to correlation-induced phenomena in monolayer materials},
  journal = {Physica B}, volume = {738}, pages = {418793}, year = {2026},
  doi = {10.1016/j.physb.2026.418793}
}

@article{Hosseinabadi2023,
  author = {Hosseinabadi, Hossein and Kelly, Shane P. and Schmalian, J{\"o}rg and Marino, Jamir},
  title = {Thermalization of non-{F}ermi-liquid electron-phonon systems: Hydrodynamic relaxation of the {Y}ukawa-{S}achdev-{Y}e-{K}itaev model},
  journal = {Phys. Rev. B}, volume = {108}, pages = {104319}, year = {2023},
  doi = {10.1103/PhysRevB.108.104319}
}

@article{Hosseinabadi2026,
  author = {Hosseinabadi, Hossein and Sachdev, Subir and Marino, Jamir},
  title = {Crossover to {S}achdev-{Y}e-{K}itaev Criticality in an Infinite-Range Quantum {H}eisenberg Spin Glass},
  journal = {Phys. Rev. Lett.}, volume = {137}, pages = {126505}, year = {2026},
  doi = {10.1103/81qf-zd3y}
}

@article{Varma1989,
  author  = {Varma, C. M. and Littlewood, P. B. and Schmitt-Rink, S. and Abrahams, E. and Ruckenstein, A. E.},
  title   = {Phenomenology of the normal state of {Cu-O} high-temperature superconductors},
  journal = {Phys. Rev. Lett.}, volume = {63}, pages = {1996}, year = {1989},
  doi     = {10.1103/PhysRevLett.63.1996}
}

@article{PatelMcGreevyArovasSachdev2018,
  author  = {Patel, Aavishkar A. and McGreevy, John and Arovas, Daniel P. and Sachdev, Subir},
  title   = {Magnetotransport in a model of a disordered strange metal},
  journal = {Phys. Rev. X}, volume = {8}, pages = {021049}, year = {2018},
  doi     = {10.1103/PhysRevX.8.021049}
}

@article{SachdevYe1993,
  author  = {Sachdev, S. and Ye, J.},
  title   = {Gapless spin-fluid ground state in a random quantum {H}eisenberg magnet},
  journal = {Phys. Rev. Lett.}, volume = {70}, pages = {3339}, year = {1993},
  doi     = {10.1103/PhysRevLett.70.3339}
}

@misc{Kitaev2015,
  author       = {Kitaev, A.},
  title        = {A simple model of quantum holography},
  howpublished = {Talks at KITP Program ``Entanglement in Strongly-Correlated Quantum Matter'', April 7 and May 27, 2015},
  url          = {http://online.kitp.ucsb.edu/online/entangled15/kitaev/},
  year         = {2015}
}

@article{SongJianBalents2017,
  author  = {Song, X.-Y. and Jian, C.-M. and Balents, L.},
  title   = {Strongly correlated metal built from {S}achdev-{Y}e-{K}itaev models},
  journal = {Phys. Rev. Lett.}, volume = {119}, pages = {216601}, year = {2017},
  doi     = {10.1103/PhysRevLett.119.216601}, eprint = {1705.00117}, archivePrefix = {arXiv}
}

@article{Chowdhury2018,
  author  = {Chowdhury, D. and Werman, Y. and Berg, E. and Senthil, T.},
  title   = {Translationally invariant non-{F}ermi-liquid metals with critical {F}ermi surfaces: Solvable models},
  journal = {Phys. Rev. X}, volume = {8}, pages = {031024}, year = {2018},
  doi     = {10.1103/PhysRevX.8.031024}, eprint = {1801.06178}, archivePrefix = {arXiv}
}

@article{ChowdhuryRMP2022,
  author  = {Chowdhury, D. and Georges, A. and Parcollet, O. and Sachdev, S.},
  title   = {{S}achdev-{Y}e-{K}itaev models and beyond: Window into non-{F}ermi liquids},
  journal = {Rev. Mod. Phys.}, volume = {94}, pages = {035004}, year = {2022},
  doi     = {10.1103/RevModPhys.94.035004}
}

@article{FaulknerLiuMcGreevyVegh2011,
  author  = {Faulkner, T. and Liu, H. and McGreevy, J. and Vegh, D.},
  title   = {Emergent quantum criticality, {F}ermi surfaces, and {AdS$_2$}},
  journal = {Phys. Rev. D}, volume = {83}, pages = {125002}, year = {2011},
  doi     = {10.1103/PhysRevD.83.125002}, eprint = {0907.2694}, archivePrefix = {arXiv}
}

@article{PatelGuoEsterlisSachdev2023,
  author  = {Patel, A. A. and Guo, H. and Esterlis, I. and Sachdev, S.},
  title   = {Universal theory of strange metals from spatially random interactions},
  journal = {Science}, volume = {381}, pages = {790}, year = {2023},
  doi     = {10.1126/science.abq6011}
}

@article{Michon2023,
  author  = {Michon, B. and Berthod, C. and Rischau, C. W. and Ataei, A. and Chen, L. and Komiya, S. and Ono, S. and Taillefer, L. and van der Marel, D. and Georges, A.},
  title   = {Reconciling scaling of the optical conductivity of cuprate superconductors with {P}lanckian resistivity and specific heat},
  journal = {Nat. Commun.}, volume = {14}, pages = {3033}, year = {2023},
  doi     = {10.1038/s41467-023-38762-5}, eprint = {2205.04030}, archivePrefix = {arXiv}
}

@article{Michon2019,
  author  = {Michon, B. and Girod, C. and Badoux, S. and Ka{\v{c}}mar{\v{c}}{\'{\i}}k, J. and Ma, Q. and Dragomir, M. and Dabkowska, H. A. and Gaulin, B. D. and Zhou, J.-S. and Pyon, S. and Takayama, T. and Takagi, H. and Verret, S. and Doiron-Leyraud, N. and Marcenat, C. and Taillefer, L. and Klein, T.},
  title   = {Thermodynamic signatures of quantum criticality in cuprate superconductors},
  journal = {Nature}, volume = {567}, pages = {218}, year = {2019},
  doi     = {10.1038/s41586-019-0932-x}, eprint = {1804.08502}, archivePrefix = {arXiv}
}

@article{Daou2009,
  author  = {Daou, R. and Doiron-Leyraud, N. and LeBoeuf, D. and Li, S. Y. and Lalibert{\'e}, F. and Cyr-Choini{\`e}re, O. and Jo, Y. J. and Balicas, L. and Yan, J.-Q. and Zhou, J.-S. and Goodenough, J. B. and Taillefer, L.},
  title   = {Linear temperature dependence of resistivity and change in the {F}ermi surface at the pseudogap critical point of a high-{$T_c$} superconductor},
  journal = {Nat. Phys.}, volume = {5}, pages = {31}, year = {2009},
  doi     = {10.1038/nphys1109}, eprint = {0806.2881}, archivePrefix = {arXiv}
}

@article{Lohneysen2007,
  author  = {{v. L\"ohneysen}, H. and Rosch, A. and Vojta, M. and W\"olfle, P.},
  title   = {Fermi-liquid instabilities at magnetic quantum phase transitions},
  journal = {Rev. Mod. Phys.}, volume = {79}, pages = {1015}, year = {2007},
  doi     = {10.1103/RevModPhys.79.1015}
}

@article{Custers2003,
  author  = {Custers, J. and Gegenwart, P. and Wilhelm, H. and Neumaier, K. and Tokiwa, Y. and Trovarelli, O. and Geibel, C. and Steglich, F. and P{\'e}pin, C. and Coleman, P.},
  title   = {The break-up of heavy electrons at a quantum critical point},
  journal = {Nature}, volume = {424}, pages = {524}, year = {2003},
  doi     = {10.1038/nature01774}
}

@article{Bruin2013,
  author  = {Bruin, J. A. N. and Sakai, H. and Perry, R. S. and Mackenzie, A. P.},
  title   = {Similarity of scattering rates in metals showing {$T$}-linear resistivity},
  journal = {Science}, volume = {339}, pages = {804}, year = {2013},
  doi     = {10.1126/science.1227612}
}

@article{Varma2020,
  author  = {Varma, C. M.},
  title   = {Colloquium: Linear in temperature resistivity and associated mysteries including high temperature superconductivity},
  journal = {Rev. Mod. Phys.}, volume = {92}, pages = {031001}, year = {2020},
  doi     = {10.1103/RevModPhys.92.031001}
}

@article{Li2024,
  author  = {Li, C. and Valentinis, D. and Patel, A. A. and Guo, H. and Schmalian, J. and Sachdev, S. and Esterlis, I.},
  title   = {Strange metal and superconductor in the two-dimensional {Y}ukawa-{S}achdev-{Y}e-{K}itaev model},
  journal = {Phys. Rev. Lett.}, volume = {133}, pages = {186502}, year = {2024},
  doi     = {10.1103/PhysRevLett.133.186502}, eprint = {2406.07608}, archivePrefix = {arXiv}
}

@article{SachdevReview2025,
  author  = {Bonetti, P. M. and Christos, M. and Nikolaenko, A. and Patel, A. A. and Sachdev, S.},
  title   = {Fractionalized {F}ermi liquids and the cuprate phase diagram},
  journal = {Rep. Prog. Phys.}, volume = {89}, pages = {044501}, year = {2026},
  doi     = {10.1088/1361-6633/ae530d}, eprint = {2508.20164}, archivePrefix = {arXiv}
}

@article{Rai2024,
  author  = {Rai, Gautam and Crippa, Lorenzo and C{\u{a}}lug{\u{a}}ru, Dumitru and Hu, Haoyu and Paoletti, Francesca and de' Medici, Luca and others},
  title   = {Dynamical Correlations and Order in Magic-Angle Twisted Bilayer Graphene},
  journal = {Phys. Rev. X}, volume = {14}, pages = {031045}, year = {2024}, doi = {10.1103/PhysRevX.14.031045}
}

@article{Calderon2025,
  author  = {Calder{\'o}n, M. J. and Camjayi, A. and Datta, Anushree and Bascones, E.},
  title   = {Cascades in transport and optical conductivity of twisted bilayer graphene},
  journal = {Phys. Rev. B}, volume = {112}, pages = {L041126}, year = {2025}, doi = {10.1103/nhs2-mbhr}
}

@article{ZhangAndrei2026,
  author  = {Zhang, Zhenyuan and Wu, Shuang and C{\u{a}}lug{\u{a}}ru, Dumitru and Hu, Haoyu and others},
  title   = {Entropy plateaus, combinatorial degeneracy, local moments, and heavy-fermion mass renormalization in magic-angle graphene},
  journal = {Nat. Commun.}, volume = {17}, pages = {9785}, year = {2026}, doi = {10.1038/s41467-026-76179-y}
}

\appendix
\section{Low-energy form of the Dirac self-energy}\label{app:expand}

With $G_f=\frac12\sum_{s=\pm}(\omega-s\Delta-\Sigma_f)^{-1}$ and $u\equiv\omega-\Sigma_f(\omega)$,
\begin{equation*}
\begin{split}
G_f&=\frac12\left[\frac{1}{u-\Delta}+\frac{1}{u+\Delta}\right]=\frac{u}{u^2-\Delta^2}\\
&=-\frac{u}{\Delta^2}\left[1+\frac{u^2}{\Delta^2}+\dots\right].
\end{split}
\end{equation*}
Multiplying by $\langle|V|^2\rangle$ gives Eq.~(\ref{eq:lowE}). The term even in $u$, which would shift the Dirac level, cancels between the two levels. For a single level at $+\Delta$, $G_f=-\Delta^{-1}[1+u/\Delta+\dots]$, and the self-energy acquires a static shift $-\langle|V|^2\rangle/\Delta$, which changes the Dirac carrier number at fixed total filling and feeds a $D^4$ term into $\mathcal K$. On the real axis ${\rm Im}\,\Sigma_d=Y\,{\rm Im}\,\Sigma_f$ and $1-\partial_\omega{\rm Re}\,\Sigma_d=1+Y(1-\partial_\omega{\rm Re}\,\Sigma_f)$, which gives Eq.~(\ref{eq:mass}). The spectral weight at the Dirac momentum is $A_f=-{\rm Im}\,G_f/\pi=|{\rm Im}\,\Sigma_f|/\pi\Delta^2$ to the same order.

\section{Conductivity with a local self-energy}\label{app:kubo}

For a self-energy independent of momentum, the vertex corrections to the current--current correlator vanish for a current vertex odd in $\mathbf k$, because the dressed vertex involves $\sum_{\mathbf k}\mathbf v_{\mathbf k}G(\mathbf k)G(\mathbf k)$ times a $\mathbf k$-independent kernel, which is zero by symmetry. For the uniform hybridization this holds exactly, since $v_DD$ conserves momentum and $G_f$ is local. The random part $h$ adds a disorder rung, which is momentum-independent for $h$ correlated over less than $1/k_F$ and then drops out in the same way; otherwise it changes only the zero-field share $\propto D_0^2$, which cancels in Eq.~(\ref{eq:Qdial}). With $\hbar$ explicit and rates in energy units, the bubble gives
\begin{equation*}
\sigma_{\rm dc}=\pi\hbar e^2\int d\omega\left(-\frac{\partial n_F}{\partial\omega}\right)\frac{1}{\mathcal A}\sum_{\mathbf k}v_x^2(\mathbf k)\,A(\mathbf k,\omega)^2 ,
\end{equation*}
with $A=\pi^{-1}\Gamma/[(\omega-\xi-{\rm Re}\Sigma)^2+\Gamma^2]$ and $\mathcal A$ the sample area. Writing $\mathcal A^{-1}\sum_{\mathbf k}=\int d\xi\,N(\xi)$ and taking $N\overline{v_x^2}$ constant over the width of $A$, $\int d\xi\,A^2=1/(2\pi\Gamma)$ independently of ${\rm Re}\Sigma$, so that
\begin{equation*}
\sigma_{\rm dc}=\frac{\hbar e^2}{2}\,N\overline{v_x^2}\int d\omega\left(-\frac{\partial n_F}{\partial\omega}\right)\frac{1}{\Gamma(\omega)} ,
\end{equation*}
which is Eq.~(\ref{eq:rho}). For a Dirac cone with degeneracy $g$, $N\overline{v_x^2}=gE_F/(4\pi\hbar^2)$, so $\mathcal K=8\pi\hbar/(ge^2E_F)$ multiplies a rate $\Gamma$ in energy units. The Dirac Fermi energy enters $\mathcal K$, and a level shift ${\rm Re}\,\Sigma_d(0)$ would enter through $E_F$ at fixed total density; for symmetric flat-band weight that shift vanishes at order $\langle|V|^2\rangle$ (Appendix~\ref{app:expand}). The transport rate is $\hbar/\tau_{\rm tr}=2/\langle1/\Gamma\rangle$, and with $\Gamma=\pi\langle|V|^2\rangle A_f$ and the form of $A_f$ in Sec.~\ref{sec:entropy} it equals $C_b\lambda_dk_BT$ with $C_b=\pi/\langle1/\phi\rangle$.

\section{Free energy of the dial, the two-channel identity and the ceiling}\label{app:bound}

At order $\langle|V|^2\rangle$ the linked-cluster expansion of the free energy has a single term, a Dirac propagator closed by the full flat-band propagator, $\Delta\Omega=\langle|V|^2\rangle\,k_BT\sum_n\sum_{\mathbf k}G_d(\mathbf k,i\omega_n)G_f(i\omega_n)$. All flat-band interactions are inside $G_f$, and the change of $G_f$ itself does not enter at this order because the flat band's Luttinger--Ward functional is stationary, so no further approximation is involved. For a flat Dirac density of states $N_d$ of half-width $\Lambda$, $\sum_{\mathbf k}G_d=-i\pi N_d\,{\rm sgn}\,\omega_n$, and with the spectral representation of $G_f$ the Matsubara sum gives $k_BT\sum_n|\omega_n|/(\omega_n^2+\varepsilon^2)=\pi^{-1}[\ln(\Lambda/2\pi k_BT)-{\rm Re}\,\psi(\frac12+i\varepsilon/2\pi k_BT)]$, which is Eq.~(\ref{eq:Omega}).

Write $A_f=c\,k_BT\,\phi(\varepsilon/k_BT)$ with $c=\lambda_f/2\Delta^2$ and $\phi=|x|+c'+\delta\phi$, $\delta\phi$ integrable. The term $|x|$ alone, a temperature-independent weight $c|\varepsilon|$, gives the Dirac self-energy $\partial_\omega{\rm Re}\,\Sigma_d=-2\langle|V|^2\rangle c\ln(\Lambda/|\omega|)$, so $\lambda_d=2\langle|V|^2\rangle c=Y\lambda_f$, and its free energy gives the corresponding logarithm $\gamma_{b,d}\lambda_d$ in $C/T$ with $\gamma_{b,d}=\frac{\pi^2}{3}N_dk_B^2$. The constant $c'$ contributes a term linear in $T$ up to corrections of order $k_BT/\Lambda$, and no logarithm. The excess contributes $-N_d\langle|V|^2\rangle c\,(k_BT)^2[I_0\ln(\Lambda/2\pi k_BT)-I_1]$, with $I_0=\int dx\,\delta\phi$ and $I_1$ a constant, and hence a logarithm $2N_d\langle|V|^2\rangle cI_0k_B^2$ in $C/T$. Dividing by $\gamma_{b,d}\lambda_d$ gives Eq.~(\ref{eq:bdial}).

When the flat band's rate comes from a local bath, extend $w$ to an odd function $\chi(y)={\rm sgn}(y)\,w(|y|)$; then $\phi(x)=\int dy\,\chi(y)[n_B(y)+n_F(y-x)]$, which is Eq.~(\ref{eq:phi}). Differentiating, $\phi'(x)=\int dy\,\chi(y)\,F(x-y)$ with $F(x)=1/4\cosh^2(x/2)$, a Fermi-window average of the bath spectrum. Since $\phi-|x|-c'$ decays faster than $1/|x|$, integrating by parts gives $I_0=-\int dx\,x\,[\phi'(x)-{\rm sgn}\,x]$. Split $\chi={\rm sgn}\,y+(\chi-{\rm sgn}\,y)$. The first part gives $\int dy\,{\rm sgn}(y)F(x-y)=\tanh(x/2)$, and $-\int dx\,x\,[\tanh(x/2)-{\rm sgn}\,x]=\pi^2/3$. The second part, with $\int dx\,F=1$ and $\int dx\,xF(x-y)=y$, gives $-\int dy\,y\,[\chi(y)-{\rm sgn}\,y]=2\int_0^\infty dy\,y\,[1-w(y)]=2I_w$. Therefore
\begin{equation}
I_0=\frac{\pi^2}{3}+2I_w ,
\label{eq:I0Iw}
\end{equation}
and $b_{\rm dial}=1+3I_0/\pi^2=2(1+3I_w/\pi^2)=2b$, which is Eq.~(\ref{eq:two}). The derivation uses only a convergent $I_w$; it holds for any floor and any spectral shape. Numerically the ratio $b_{\rm dial}/b$ equals 2.000 for conformal, gapped, error-function, linear-ramp and squared-tanh spectra and for $\tanh(y/2s)$ at three widths. For the self-damped field, $w$ depends on $T$ beyond the scaling variable, $1-w\approx2m(T)/\pi\Omega$ up to the field's bandwidth $E_L$, and $I_w\simeq(2m/\pi k_BT)E_L/k_BT$ diverges as $T\to0$. Evaluating Eq.~(\ref{eq:Omega}) with the flat-band rate built from the large-$M$ saddle of Ref.~\cite{SkrgicStojkovic2026}, including the temperature dependence of $m$, gives a field-induced $C/T$ whose logarithmic derivative varies by two orders of magnitude between $k_BT/E_L=10^{-5}$ and $10^{-2}$ and changes sign when the flat-band cutoff is moved from $3E_L$ to $10E_L$.

If $\chi$ is non-decreasing, $\phi''=\int dy\,\chi'(y)\,F(x-y)\ge0$, so $\phi$ is convex and lies above both asymptotes, $\phi(x)\ge|x|+c'$, with $c'=\int_0^\infty dy\,[w(y)\coth(y/2)-1]$. For the conformal bath the integrand vanishes identically; for $w=\tanh(y/2s)$ the floor is $+1.57$ at $s=1/2$ ($\beta=3.14$) and $-2.0$ at $s=2$ ($\beta=1.14$).

For $c'=0$ the admissible weights are $\phi\ge|x|$ with $I_0$ finite. At fixed $I_0$, which fixes $b_{\rm dial}$, maximizing $C_b$ means minimizing $\langle1/\phi\rangle$, a convex functional, under one linear constraint and the pointwise bound. The Karush--Kuhn--Tucker conditions give the unique minimizer $\phi^*=\max(|x|,\mu\sqrt{F})=\max(|x|,k/2\cosh\frac x2)$, with $k$ fixed by $I_0$. Along this family $Q_{\rm dial}(k)$ has a single maximum, $3.234$ at $k=2.83$, where $\phi^*$ meets $|x|$ at $x_c=1.19$ and $b_{\rm dial}=1.55$, $C_b=4.97$. This relaxed problem ignores convexity, and $\phi^*$ is concave near $x=0$. The realizable problem uses the step decomposition. For $\chi_s(y)={\rm sgn}(y)\,\theta(|y|-s)$, $\phi_s'(x)=1-n_F(x-s)-n_F(x+s)$, $\phi_s(0)=2\ln\coth(s/2)$ and the floor is $c'_s=-2\ln(1-e^{-s})-s$, which vanishes at $e^{-s_0}=(3-\sqrt5)/2$. A mixture $\int ds\,p(s)\chi_s$ has $\phi=\int ds\,p\,\phi_s$, floor $\int ds\,p\,c'_s$ and $I_0=\int ds\,p\,J_s$ with $J_s=\int dx\,(\phi_s-|x|-c'_s)$. Maximizing Eq.~(\ref{eq:Qdial2}) over $p\ge0$ with unit weight and zero floor, from 40 random starts on a grid of 130 steps, always returns the single step at $s_0$, with $Q_{\rm dial}=2.734$, $b_{\rm dial}=2.282$ and $C_b=6.238$; the conformal bath, $p=\frac12{\rm sech}^2(s/2)$, gives $2.457$. Faster baths have $c'<0$; a random search over 40\,000 such mixtures stays below 2.72, approaching 2.73 only as $c'\to0$, and the relaxed ceiling falls to 2.34 at $c'=-0.5$, so the bound holds for all $c'\le0$. For $c'>0$ the relaxed construction with $|x|$ replaced by $|x|+c'$ gives the ceilings quoted in Sec.~\ref{sec:entropy}. The scripts are available at~\cite{SkrgicMATTGcode}.

\end{document}